\documentclass[12pt]{article}
\usepackage{epsfig,cite}

\begin{document}
\setlength{\unitlength}{0.2cm}

\title{
Bilocal Dynamics for Self-Avoiding Walks
}
\author{
  \\
  {\rm Sergio Caracciolo$^{\rm a}$, Maria Serena Causo$^{\rm b}$,
          Giovanni Ferraro$^{\rm c}$,} \\
  {\rm  Mauro Papinutto$^{\rm d}$, and
          Andrea Pelissetto$^{\rm c}$}             \\[0.2cm]
  {\small\it ${}^{\rm a}$ 
      Scuola Normale Superiore and INFN -- Sez. di Pisa, I-56100 Pisa, ITALY} 
        \\[-0.2cm]
  {\small\it ${}^{\rm b}$ 
      NIC, Forschungszentrum J\"ulich, D-52425 J\"ulich, GERMANY}   
        \\[-0.2cm]
  {\small\it ${}^{\rm c}$ 
      Dip. di Fisica and INFN -- Sez. di Roma I, 
      Universit\`a di Roma I, I-00185 Roma, ITALY} 
        \\[-0.2cm]
  {\small\it ${}^{\rm d}$ 
      Dip. di Fisica and INFN -- Sez. di Pisa, 
      Universit\`a di Pisa, I-56100 Pisa, ITALY} 
        \\[0.2cm]
{\small E-mail: {\tt Sergio.Caracciolo@sns.it}, {\tt M.S.Causo@fz-juelich.de},}
   \\[-0.2cm]
{\small
                  {\tt papinutt@cibs.sns.it},  
                  {\tt Andrea.Pelissetto@roma1.infn.it} }\\
}
\vspace{0.5cm}

\maketitle
\thispagestyle{empty}   

\vspace{0.2cm}

\begin{abstract}
We introduce several bilocal algorithms for lattice self-avoiding walks that 
provide reasonable models for the physical kinetics of polymers 
in the absence of hydrodynamic effects.
We discuss their ergodicity in different confined
geometries, for instance in strips and in slabs. 
A short discussion of the dynamical properties in the absence of 
interactions is given.
\end{abstract}

\clearpage

\newcommand{\be}{\begin{equation}}
\newcommand{\ee}{\end{equation}}
\newcommand{\beq}{\begin{equation}}
\newcommand{\eeq}{\end{equation}}
\newcommand{\bea}{\begin{eqnarray}}
\newcommand{\eea}{\end{eqnarray}}
\newcommand{\<}{\langle}
\renewcommand{\>}{\rangle}

\def\spose#1{\hbox to 0pt{#1\hss}}
\def\ltapprox{\mathrel{\spose{\lower 3pt\hbox{$\mathchar"218$}}
 \raise 2.0pt\hbox{$\mathchar"13C$}}}
\def\gtapprox{\mathrel{\spose{\lower 3pt\hbox{$\mathchar"218$}}
 \raise 2.0pt\hbox{$\mathchar"13E$}}}

\newcommand{\R}{\hbox{{\rm I}\kern-.2em\hbox{\rm R}}}
\newcommand{\reff}[1]{(\ref{#1})}
\def\smfrac#1#2{{\textstyle\frac{#1}{#2}}}

\section{Introduction}

The lattice self-avoiding walk (SAW) is a well-known model for the 
critical behaviour of a homopolymer in a 
solvent \cite{deGennes_book,desCloizeaux-Jannink_book} 
and it has been extensively used in the study of various 
properties of heteropolymers \cite{Lau-Dill_89,Sali-etal_94}. 
Experiments are usually performed using monodisperse solutions and thus 
extensive work has been
done to devise Monte Carlo algorithms to simulate fixed-length SAWs. 
Historically, the earliest algorithms used a local dynamics
\cite{Verdier-Stockmayer_62}: at each step a small part of the walk
(usually 2-4 consecutive beads) was modified. Although easy to implement,
these algorithms suffer a very serious drawback: as shown by 
Madras and Sokal \cite{Madras-Sokal_87},
any local algorithm is not ergodic and simulations span only an 
exponentially small part of the phase space. A different algorithm
was inspired by an attempt to model the true dynamics of the polymer in
the solvent: the reptation algorithm \cite{Kron_65,Kron-etal_67,%
Wall-Mandel_75,Mandel_79}. However, it was soon realized 
\cite{Kron-etal_67,Wall-Mandel_75}
that it is not ergodic because of the possibility of 
configurations with trapped endpoints. These ergodicity problems 
can be solved using chain-growth algorithms 
\cite{Suzuki_68,Redner-Reynolds_81,Grassberger_97}
or non-local algorithms
\cite{Lal_69,MacDonald-etal_85,Madras-Sokal_88,Madras-etal_90,%
Caracciolo-etal_92}.
In the absence of any interaction, non-local algorithms are 
very efficient. For instance, in the pivot algorithm 
\cite{Madras-Sokal_88} the autocorrelation
time for global observables increases linearly with the number of steps
$N$, which is the optimal behaviour since it takes a time of order
$N$ simply to write down the walk. 

Non-local algorithms are extremely efficient in the absence of interactions
and in free space. However, in the presence of strong attraction 
and in finite geometries non-local moves are 
largely inefficient\footnote{An exception is the class of moves 
introduced in \cite{Deutsch_97} that do not change the position of the 
beads but only the connectivity of the walk. However these moves do
not change global size observables and are of interest only
for the study of maximally compact configurations.
A general lower bound on the efficiency of non-local algorithms with 
a Metropolis test is given in Ref. \cite{Caracciolo-etal_94}.}.
Indeed, if the interactions are strongly attractive
(for instance at the $\theta$-point or in the collapsed phase),
typical configurations are compact so that the probability of success
of non-local moves is very small. Therefore, the  algorithm 
becomes inefficient. In the presence of surfaces, 
non-local algorithms are not even ergodic in general. For instance,
as we will show,  
the pivot algorithm is not ergodic in a strip. Moreover, even
when they are ergodic, they are not suited to study surface transitions
since non-local moves will generate new walks with large energy
differences and thus they will be rejected. Again the dynamics will be 
very slow.

Stochastic algorithms are also important for the study of the 
physical kinetics of polymers. 
There is now a widespread interest in understanding the 
dynamics of homopolymers in the collapsed phase 
\cite{Chan-Dill_93}
and of heteropolymers near the folding temperature
\cite{Karplus-Shakhnovich_book,Bryngelson-etal_95,Grassberger_book}. In these 
studies one should use a dynamics which resembles the true dynamics 
of the molecule in a solvent. Local algorithms
cannot provide good physical models because, as we said above, 
they are not ergodic and because, as noted in Ref. 
\cite{Skolnick-Kolinski_91}, they do not allow diffusive motions 
along the chain: for instance, they cannot move around assembled
pieces of secondary structure\footnote{In spite of these 
problems a local dynamics is of widespread use,
see e.g. \cite{Chan-Dill_94,Sali-etal_94b,Socci-Onuchic_94,%
Nunes-etal_96,Gutin-etal_98,Cieplak-etal_98,Collet_99}. Ref. 
\cite{Socci-Onuchic_94} claims that, at the presently investigated values 
of $N$, non-ergodicity effects should be small. A quantitative 
study of the Verdier-Stockmayer algorithm in two dimensions 
\cite{Verdier-Stockmayer_62}
was presented in Ref.  \cite{Madras-Sokal_87}.  The percentage of walks 
that do not belong to the ergodicity class of the straight rod is 
indeed small, precisely 
0.0067\%, 0.0061\%, 0.0041\% for $N=11$, 13, 15. However, 
it should be noted 
that this percentage increases with $N$ (each ergodicity class 
contains only an exponentially small fraction of the walks
\cite{Madras-Sokal_87}) and 
that the walks that do not belong to the ergodicity class 
of the straight rod
correspond to compact configurations: therefore, larger systematic
deviations are expected in the collapsed regime. 
If we indicate with $c_N(n)$ the number of walks with $N$ steps 
and $n$ nearest-neighbor contacts and with $d_N(n)$ 
the corresponding number of walks that do not belong to the ergodicity class
of the straight rod, a good indication of the deviations expected 
in the presence of strong attraction is given by 
$R_N = d_N(n_{max})/c_N(n_{max})$ and 
$S_N(\epsilon) = \sum_n e^{n\epsilon} d_N(n)/\sum_n e^{n\epsilon} c_N(n)$.
Here $n_{max}$ is the maximum number of possible contacts for a given 
$N$; clearly $R_N = S_N(\infty)$. For the Verdier-Stockmayer algorithm
we considered above, we have $R_N =$ 3.2\%, 1.4\%, 5.8\% for $N=11$, 13, 15,
and $S_N(1) =$ 0.20\%, 0.20\%, 0.16\% for the same values of $N$. 
Clearly, the systematic error is not completely negligible in the compact 
regime.
}... 
As we said above, the ergodicity problem can be solved by using
non-local algorithms,
and indeed, non-local moves have been considered in
\cite{Skolnick-Kolinski_91,Kolinski-Skolnick_94,Covell_94}. 
However a non-local dynamics which involves rigid deformations of a large
section of the polymer is unphysical, and therefore cannot give 
realistic results for the physical kinetics.

In this paper we wish to discuss a family of algorithms that use 
bilocal moves: a bilocal move alters at the same time two
disjoint small groups of consecutive sites of the walk 
that may be very far away. 
Since a small number of beads is changed at each step, these algorithms
should be reasonably efficient in the presence of interactions, 
and thus they can be used in the study of the collapsed phase 
and of the folding of heteropolymers. They generalize 
the reptation algorithm and use a more general class of moves 
that was introduced by Reiter \cite{Reiter_90}. 
Similar moves were introduced in Ref. \cite{Jagodzinski-etal_92}
and were applied to the study of ring polymers\footnote{However, 
it should be noted that the algorithm of Ref. \cite{Jagodzinski-etal_92} 
is not ergodic.}. 

We will study in detail the ergodicity of these algorithms and we will show 
that, with a proper choice of moves, they are ergodic even
in some constrained geometries, e.g. in strips, slabs, and generalizations 
thereof. These results have been obtained for SAWs with nearest-neighbour
jumps on a (hyper-)cubic lattice. However, they can be easily
generalized to different lattice models, for instance to the 
protein model proposed in Ref. \cite{Kolinski-etal_91}.
Since these algorithms change a small number of beads at each step, they can 
be used to study the physical dynamics
of dilute polymers in the absence of hydrodynamic interactions 
and are therefore a suitable generalization of the Rouse model
\cite{Rouse_53,Doi_book}
in the presence of excluded-volume effects. Therefore 
they provide a good model for the collapse of homopolymers and for the 
folding dynamics\footnote{Note that in a bilocal move a large piece of 
the walk moves of one or two lattice steps. In protein models with 
nearest-neighbour interactions this may generate large energy differences.
In this case, it is probably important to consider interactions with a longer 
range. This corresponds to considering microscopic motions of the chain
that are small compared to the interaction range. Of course, these 
problems do not arise in continuum models, in which one can make steps as 
small as one wishes.}. 

The paper is organized as follows. 
In Sect. 2 we introduce several local and bilocal moves and we define three 
bilocal algorithms. In Sect. 3 we discuss their ergodicity,
determining the minimal set of moves that make each algorithm ergodic.
This is important in order to understand the dynamical behaviour.
Indeed, algorithms that are ergodic only if rarely accepted or rarely proposed
moves are included, are expected {\em a priori} to have a slow 
dynamics. Sect. 4 contains a detailed presentation of their
implementation. In the last Section we present a brief discussion
of the expected dynamic critical behaviour in the absence of interactions. 
A detailed numerical study will appear elsewhere\cite{Caracciolo-etal_2000}.

\section{Definition of the algorithms}

In this paper we will consider SAWs with fixed number of steps $N$ 
and free endpoints in finite geometries.

More precisely, we consider a $d$-dimensional hyper-cubic lattice and 
define the following set of lattice points:
given an integer $D$ such that $1\le D \le d-1$, and $(d-D)$ positive integers
$w_{D+1},\ldots, w_d$, we define 
${\cal C}_D(w_{D+1},\ldots, w_d)$ as the set of lattice points 
$(n_1,\ldots,n_d)$, $n_i\in Z$, such that 
$0\le n_i \le w_i$, for $i=D+1,\ldots,d$. We will call\footnote{
In two dimensions one often calls cylinder a strip with periodic boundary 
conditions. Note that the definition given  here is different.}
${\cal C}_D(w_{D+1},\ldots, w_d)$
a $D$-dimensional {\em cylinder}. If $D=(d-1)$, we will speak 
of a strip if $d=2$ and of a slab if $d=3$. The number $w_i$ will be called 
the {\em width} of the cylinder in the $i$-th direction. 
Note that we will always assume $D\ge 1$, so that at least the first 
direction is infinite. 

We will then consider SAWs of length $N$ confined inside a cylinder. 
A SAW $\omega$ is a set 
of $N+1$ lattice points $\omega(0)$, $\ldots$, 
$\omega(N)$ such that: $\omega(i)$ and $\omega(i+1)$ are lattice 
nearest neighbours; $\omega(i)\not=\omega(j)$ for any $i\not=j$;
$\omega(i)\in {\cal C}_D(w_{D+1},\ldots, w_d)$ for all $i$. We will 
define two different ensembles:
\begin{enumerate}
\item the ensemble ${\cal E}_{x,N}$ of SAWs of length $N$ such that 
     $\omega(0) = x$;
\item the ensemble ${\cal E}_{N}$ of SAWs of length $N$ such that both
     endpoints can be anywhere in the cylinder.
\end{enumerate}
In free space the two ensembles are equivalent as long as one is interested 
in properties of the walk itself. In confined geometries they are different
since translation invariance is lost in $d-D$ directions. Both ensembles
are of physical interest: in a slab one can study the statistical properties
of polymers that can move freely between the confining surfaces, 
or one can determine the behaviour of polymers grafted at one of the 
boundaries.

We wish now to define some algorithms that sample these ensembles
of walks. They use local and bilocal moves.
A local move is one that 
alters only a few consecutive beads of the SAW, leaving the other sites 
unchanged. A bilocal move is instead one that alters two disjoint small groups 
of consecutive sites of the walk; these two groups may in general be very
far from each other. 

\begin{figure}
\centering
\epsfig{figure=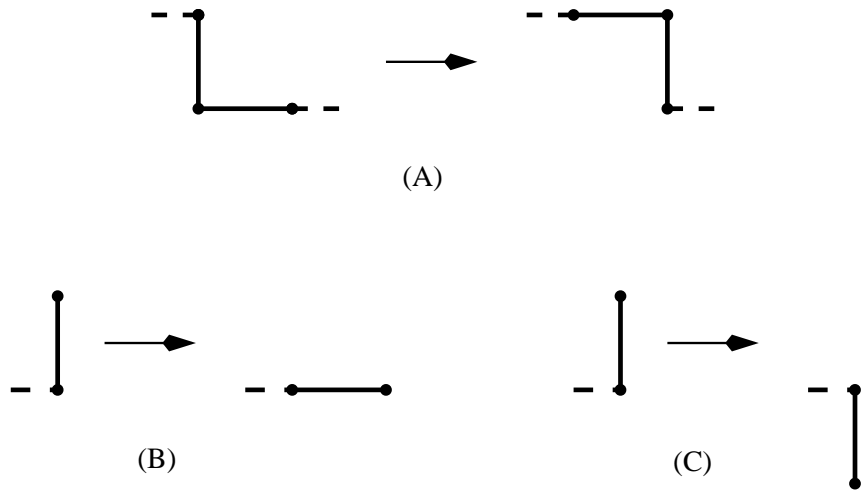,angle=-90,width=0.7\linewidth}
\vspace{0.5cm}
\caption{All one-bead moves. (A) One-bead flip. (B) $90^{\circ}$
end-bond rotation. (C) $180^{\circ}$ end-bond rotation.}
\label{one-bead}
\end{figure}

\begin{figure}
\begin{center}
\epsfxsize = 0.8\textwidth
\leavevmode\epsffile{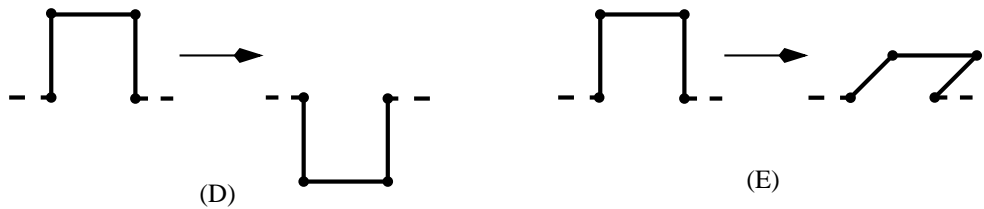}
\end{center}
\vspace*{0cm}
\caption{Kink rotations or crankshaft moves.
}
\label{two-bead}
\end{figure}

In our study we will introduce three types of local moves 
(see Figs. \ref{one-bead}, \ref{two-bead}):
\begin{itemize}
\item[{[L0]\hphantom{0}}] One-bead flips in which one bead only is moved.
\item[{[L00]}] Kink rotations (also called crankshaft moves) in which a 
  three-step kink is rotated.
\item[{[L1]\hphantom{0}}] End-bond rotations in which the last step of the 
walk is rotated. The same move can also be applied to the first step of 
the walk.
\end{itemize}

\begin{figure}
\centering
\epsfig{figure=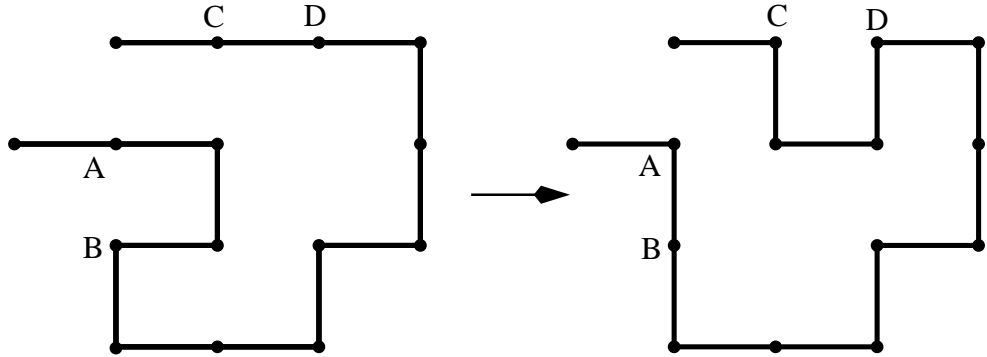,angle=-90,width=0.8\linewidth}
\vspace{0.5cm}
\caption{The kink-transport move. A kink has been cleaved from AB and
attached at CD. Note that the new kink is permitted to occupy one or
both of the sites abandoned by the old kink.}
\label{kink-trans}
\end{figure}

\begin{figure}
\centering
\vspace{0.2cm}
\epsfig{figure=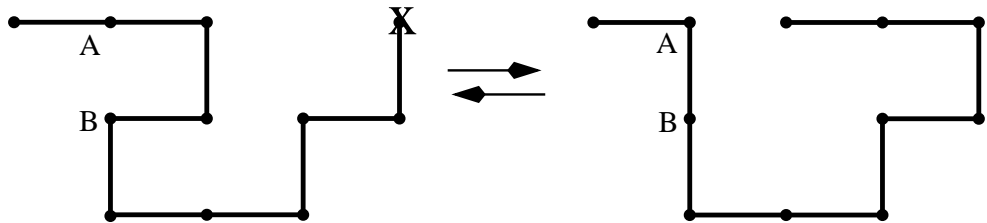,angle=-90,width=0.8\linewidth}
\vspace{0.5cm}
\caption{The kink-end reptation $(\longrightarrow)$ and end-kink
reptation $(\longleftarrow)$ moves. In $(\longrightarrow)$, a kink has
been cleaved from AB and two new steps have been attached at the end
marked X. Note that the new end steps are permitted to occupy one or
both of the sites abandoned by the kink.}
\label{kink-end}
\end{figure}

\begin{figure}
\centering
\epsfig{figure=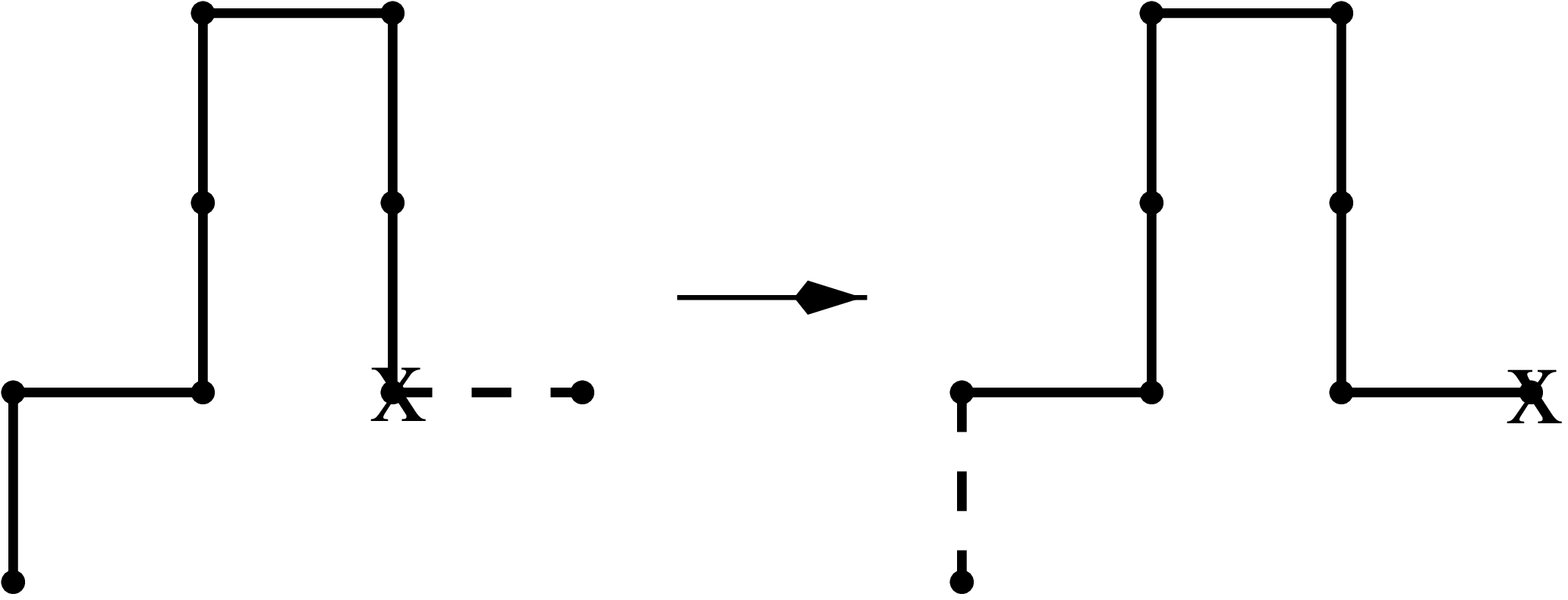,angle=-90,width=0.7\linewidth}
\vspace{0.5cm}
\caption{The reptation move. The head of the walk is indicated by
X. The dashed lines indicate the proposed new step and the 
abandoned old step.}
\label{reptation}
\end{figure}

We will then introduce several types of bilocal moves:
\begin{itemize}
\item[{[B22]}] Kink-transport moves in which a kink is cleaved from
the walk and attached at two other different sites (see Fig. \ref{kink-trans});
note that the new kink is allowed to occupy one or both of the sites 
abandoned by the old kink.
\item[{[BKE]}] Kink-end and end-kink reptation moves (see Fig. \ref{kink-end}).
In the kink-end reptation move a kink is deleted at one location along the 
walk and two new bonds are appended in arbitrary directions at the free 
endpoint of the walk. Viceversa, an end-kink reptation move consists in 
deleting two bonds from the end of the walk and in inserting a kink, 
in arbitrary orientation, at some location along the walk.
The same move can also be applied to the first step of the walk.
\item[{[BEE]}] Reptation move (see Fig. \ref{reptation}) in which one bond is 
deleted from one end of the walk and a new bond is appended in arbitrary 
direction at the other end.
\end{itemize}

We wish now to define algorithms made up with the moves we have 
presented above and that are ergodic. As shown by Madras and Sokal 
\cite{Madras-Sokal_87},
there exists no ergodic algorithm made up of local moves. It is therefore 
necessary to add some bilocal moves to obtain ergodicity. 
The oldest bilocal algorithm is the reptation algorithm, which uses 
only the moves BEE. As it was soon realized, it is not ergodic due to 
the presence of walks with trapped endpoints. Here we wish to define 
new bilocal algorithms. We will first consider the 
ensemble ${\cal E}_{x,N}$. We will discuss two algorithms:
\begin{itemize} 
\item {\em Kink-kink bilocal algorithm}. It uses the 
   local moves L0, L1, and the bilocal moves B22. 
   The local move L1 is applied only to the last point of the walk,
   otherwise $\omega(0)$ would not be kept fixed. We will show that 
   it is ergodic in two dimensions (under some technical conditions), and that 
   it is not ergodic in three dimensions due to the possibility 
   of knots. In higher dimensions its ergodicity is an open problem.
\item {\em Kink-end reptation}. It uses the moves BKE applied to the 
   last step of the walk. We will show that it is ergodic in a 
   $D$-dimensional cylinder for $d\ge 3$. In two dimensions it is 
   ergodic only in free space or in the presence of a single 
   surface, i.e. for $D=1$ and $w_2=\infty$. There exists an 
   extension that is ergodic in a two-dimensional strip: 
   it uses the moves BKE applied to the last step of the walk
   and the local moves L0.
\end{itemize}
It is trivial to modify these algorithms so that the first point is 
not kept fixed. It is enough to apply L1 and BKE moves to the first
step of the walk, too. However, these modifications are not ergodic 
in the ensemble ${\cal E}_N$: indeed, L1 and BKE moves never change the 
parity of the first point. If there is translation invariance in one 
infinite direction, this limitation is irrelevant: all walk properties 
can still be obtained correctly from the ensemble of walks in 
which the parity of $\omega(0)$ is fixed. However, this is not the case 
in the presence of random interactions, since translation invariance is 
completely lost. In order to sample the ensemble ${\cal E}_N$ we introduce
a different algorithm:
\begin{itemize}
\item {\em Extended reptation}. It uses the moves L0, B22, BEE. 
   We will show that it is ergodic in two dimensions. 
   The ergodicity for $d\ge 3$ is an open problem. 
   In the absence of a definite result, 
   an ergodic extension in $d=3$ can be obtained by adding BKE moves.
\end{itemize}

The ergodicity properties of these algorithms are proved in the next Section.
It should be noticed that we have not considered the local moves L00, which 
are not necessary for ergodicity, but that can be added to the moves L0 if one 
wishes. 

The bilocal moves we have introduced above have already been discussed in
the literature. Removals and insertions of kinks were introduced in
\cite{Berg-Foerster_81,Aragao-Caracciolo_83,Aragao-etal_83} 
in order to study cyclic 
polymers and polymers with fixed endpoints with varying length $N$.
An ergodic algorithm was introduced by Reiter \cite{Reiter_90}:
he considers moves L0, L1, B22, BEE, BKE and proves the ergodicity 
of the dynamics (his ergodicity proof requires only BKE and BEE moves)
in free space. A similar algorithm was used in \cite{Jagodzinski-etal_92}
in a study of cyclic polymers, considering L0, L00, and BEE moves. 
In two dimensions the algorithm is ergodic even in a strip, as we 
shall show below, while in three dimensions it is not ergodic since it 
does not change the knot type of the loop. It is unknown
if it is ergodic in a given knot class.

\section{Ergodicity}

\subsection{Definitions}

In this Section we will prove various ergodicity theorems for the algorithms 
we have introduced before. 

Let us begin by introducing some definitions, 
following Ref. \cite{Madras-Slade_book}:

\vskip 0.4 truecm

\noindent
{\em Definition 1}: A subwalk $\omega[i,j]$, $(0\leq i<j\leq N)$ is a
{\em C-turn} of $\omega$ if $j-i\geq 3$, $\omega[i+1,j-1]$ lies on a straight
line that is perpendicular to the steps $\Delta\omega(i)$ and
$\Delta\omega(j-1)$ and $\Delta\omega(i)=-\Delta\omega(j-1)$. \par
\noindent The length of the C-turn is the length of the subwalk. \par
\noindent
We say that a C-turn {\em belongs } to a line (or surface), or that 
a line {\em contains} a C-turn, if the segment
$\omega[i+1,j-1]$ lies on the line (or surface). 

\vskip 0.4 truecm

\noindent
{\em Definition 2}: A C-turn of $\omega$, $\omega[i,j]$ is {\em obstructed}
if there is a site of $\omega$ lying on the line segment whose endpoints are
$\omega(i)$ and $\omega(j)$. Otherwise it is {unobstructed}.

\vskip 0.4 truecm

\noindent
{\em Definition 3}: The {\em enveloping hyper-rectangle}
$R[\omega]$ of the walk $\omega$
is the lattice hyper-rectangle of minimal volume (which may be zero) 
containing $\omega$. A lattice hyper-rectangle is the set of points 
$(x_1,\ldots x_d)$, such that $l_1\le x_1 \le L_1$, $l_2\le x_2 \le L_2$,
$\ldots$, $l_d\le x_d \le L_d$, for some integers
$l_1,l_2,\ldots$, $L_1,L_2\ldots$

\vskip 0.4 truecm

\noindent
{\em Definition 4}: A walk $\omega$ is {\em directed} if there are no steps 
that have opposite directions.

\vskip 0.4 truecm

\noindent
{\em Definition 5}: A {\em tower} of links of height $h\geq0$ is a subwalk
$\omega[i,j]$ with $j-i=2h+1$, $0\leq i<j\leq N$, such that $\omega[i,i+h]$ and
$\omega[i+1+h,j]$ are segments and $\omega(i)$ and $\omega(j)$ are
lattice nearest neighbours.\par
\noindent We call the lattice link $l$ connecting $\omega(i)$ and $\omega(j)$
the {\em base} of the tower. We denote the tower as $T(l,h)$.\par
\noindent Moreover we will say that a tower is parallel to a given line if the
segments $\omega[i,i+h]$ and $\omega[i+1+h,j]$ are parallel to this line.

\vskip 0.4 truecm

\noindent
{\em Definition 6}: Given a walk $\omega$ and a tower $T(l,h)$, 
$l$ connecting $\omega(i)$ and $\omega(j)$,  we define the
{\em quotient walk}  $\omega/T(l,h)$ as the walk with sites $\omega(0)$ ....
$\omega(i)$, $\omega(j)$, .... $\omega(N)$.

\vskip 0.4 truecm

\noindent
{\em Definition 7}: Given a walk $\omega$ with a tower $T(l,h)$, let
$\overline\omega = \omega / T(l,h)$ be the quotient walk.
If $ \overline\omega $ is directed,
$ \omega $ is said to be {\em quotient-directed}.

\vskip 0.4 truecm

For the two dimensional proofs we will make extensive use of the
following theorem due to Madras and reported in \cite{Madras-Slade_book},
Theorem 9.7.2, pag. 356:

\vskip 0.4 truecm

\noindent
{\em Theorem 1}: In two dimensions, 
if a walk $\omega$ has at least one C-turn,
then $\omega$ has an unobstructed C-turn.

\vskip 0.4truecm

\subsection{Ergodicity properties of the pivot algorithm in a confined
domain}

In this Section we want to discuss the ergodicity properties 
of the pivot algorithm. Following Ref. \cite{Madras-Sokal_88} it is possible 
to prove that the algorithm is ergodic in the presence of a single 
confining surface. More precisely, the following theorem holds:

\vskip 0.4 truecm

\noindent
{\em Theorem 2}: In $d$ dimensions, 
the pivot algorithm is ergodic in a $(d-1)$-dimensional 
cylinder if
$N<\max(\omega_d(0),w_d-\omega_d(0))$, where $\omega_d(0)$ is the 
$d$-th component of $\omega(0)$ and $N$ is the number of steps 
of the walk. 

\vskip 0.4truecm

However, the algorithm is not expected to be ergodic in more constrained 
geometries.
We will now show that the pivot algorithm is 
not ergodic in a two-dimensional strip of width $w$, for $N > (w + 1)^2 $. 
Indeed, pick a bead $\omega(i)$ and let $d$ be its distance from
the boundary $y=0$. Then, consider 
the reflections with respect to the diagonals and the $\pm 90{}^0$ 
rotations. These are the only transformations that change the number 
of links that are oriented in the $\pm x$ and in the $\pm y$ directions. 
It is easy to see that these moves are successful only if either 
$-d \le \omega(j)_x - \omega(i)_x\le w - d$ or 
$d - w \le \omega(j)_x - \omega(i)_x \le d$ for all $j> i$. 
But this cannot be verified 
if $N-i > (w + 1)^2 $. Consider now the subwalk $\Omega = 
\omega[0,N-(w + 1)^2]$. The previous argument shows that the number of links
belonging to $\Omega$ that are directed in the $\pm y$ or in the 
$\pm x$ is fixed. Thus, the algorithm is not ergodic.
We believe, although we have not been able to prove, that the algorithm 
is also not ergodic in a three-dimensional slab.

\subsection{Ergodicity of the kink-kink bilocal algorithm}

We will now prove the ergodicity of the kink-kink bilocal algorithm 
in two dimensions in a strip of width $w \equiv w_2$. To simplify writing 
the walks, we indicate by N and S the positive and negative $y$-direction, 
and by W and E the positive and negative $x$-direction.
Let us begin by proving the following lemmas. In all cases we assume 
$d = 2$.

\vskip 0.4 truecm

\noindent
{\em Lemma 1}: Consider a directed walk $\omega$ and suppose
that the distance between $ \omega (0) $ and at least one boundary of the
strip is larger than or equal to 2. 
Then $\omega$ can be reduced to any
given rod using the kink-kink bilocal algorithm.

\vskip 0.1truecm

\noindent Proof :
Using L0 and L1 moves, it is trivial to show that it is possible to 
reduce the walk either to $W^N$ or to $E^N$. We will now show that it is 
possible to deform one rod into the other. If $N\le 2$, the procedure is 
trivial and thus we will assume $N\ge 3$.

Using L0 and L1 moves, we can deform $E^N$ as follows:
\be
E^N \to E^{N-1} N \to N E^{N-1} \to N E^{N-2} N \to
        N^2 E^{N-2}
\ee
where we have assumed that the distance between $ \omega (0) $ and
the upper boundary of the strip is at least 2. If this is not the case, by
hypothesis, the distance between $ \omega (0) $ and the lower
boundary of the strip is at least 2, so that the
rod $ E^N $ can be analogously reduced to $ S^2 E^{N-2} $. The steps 
we will present below should then be changed replacing N by S.
Then, by repeatedly performing the following sequence $(p\ge 2)$
\begin{eqnarray}
W^k N^2 E^p &\to &W^k N^2 E^{p-1} S \to W^k N^2 E S E^{p-2} \to
W^{k+1} N E^p \to  \nonumber \\
&& W^{k+1} N E^{p-1} N \to W^{k+1} N^2 E^{p-1},
\end{eqnarray}
we deform the walk into $W^{N-3} N^2 E$. Finally 
\be
W^{N-3} N^2 E \to W^{N-3} N^2 W \to W^{N-2} N^2 \to W^{N}.
\ee
Q.E.D.

\vskip 0.4 truecm

The proof of this lemma requires a technical hypothesis, which however 
is relevant only if $w\le 2$. For larger strips it is always satisfied.
In the following we will also assume that $\omega (0) $ is not a 
nearest neighbour of a boundary of the strip. Thus, the 
hypothesis of the lemma is satisfied also for $w = 2$.

\vskip 0.4 truecm

\noindent
{\em Lemma 2}: Consider a quotient-directed walk $\omega$ in a 
two-dimensional strip of width $w \ge 2$ and 
suppose that $ \omega (0) $ is not
a nearest neighbour of a boundary of the strip.
Then $\omega $ can be reduced to any given rod using the 
kink-kink bilocal algorithm.

\vskip 0.1truecm
\noindent Proof :
Consider $\overline\omega = \omega/T(l,h)$, and assume that all steps 
of $\overline\omega$ are directed in the $N$, $E$ directions. 
It is immediate to verify that, by using L0 and L1 moves, one can 
modify the walk obtaining one of these four possibilities:
\begin{eqnarray}
{\rm (a)} && N^{k_1} E^{k_2} N^h E S^h E^{k_3};   \nonumber \\
{\rm (b)} && N^{k_1} E^{k_2} S^h E N^h E^{k_3};   \nonumber \\
{\rm (c)} && E^{k_1} N^{k_2} E^h N W^h N^p E^{k_3}; \nonumber \\
{\rm (d)} && E^{k_1} N^{k_2} W^h N E^h N^p E^{k_3}.   \nonumber
\end{eqnarray}
Of course we assume $h>0$; otherwise the walk is already directed and it
can be reduced to any rod by lemma 1.
In case (a), we can use L0 moves to modify the walk into 
$N^{k_1 + h} E^{k_2 + k_3 + 1} S^h$ and then, combining L0 and L1 moves,
into $N^{k_1 + h} E^{h + k_2 + k_3 + 1}$.
The new walk is directed and, by lemma 1, it can be reduced to any given rod.

In case (b), using L0 and L1 moves, we modify the walk as follows:
\begin{eqnarray}
\omega &\to& N^{k_1} E^{k_2} S^h E^{k_3+1} N^h \to 
           N^{k_1} E^{k_2} S^h E^{k_3+1} N^{h-1} E \to
           N^{k_1} E^{k_2} S^h E^{k_3+2} N^{h-1} \to 
\nonumber \\
&&  N^{k_1} E^{k_2} S^h E^{h + k_3+ 1} \to \ldots \to
    N^{k_1} E^{h + k_2 + k_3 + 1} S^h \to 
    N^{k_1} E^{h + k_2 + k_3 + 1} S^{h-1} E \to  \ldots \to
\nonumber \\ 
&&  N^{k_1} E^{h + k_2 + k_3 + 2} S^{h-1} \to
    \ldots \to
    N^{k_1} E^{2 h + k_2 + k_3 + 1},
\end{eqnarray}
which is directed. By lemma 1, it can be reduced to any given rod.

Let us now consider case (c). If $p>1$, using L0 and L1 moves, 
we can deform the walk into a 
new one with $p=1$. If $p=1$ and $k_3>0$, 
using B22 moves we obtain
\be
E^{k_1} N^{k_2} E^h N W^h N E^{k_3} \to
E^{k_1} N^{k_2} E^{h+1} N W^h N E^{k_3-1} \to
\ldots \to
E^{k_1} N^{k_2} E^{h+k_3} N W^h N. 
\ee
and then, using 
\begin{eqnarray}
&& E^{k_1} N^{k_2} E^{h+k_3} N W^h N \to 
E^{k_1} N^{k_2} E^{h+k_3} N W^{h+1} \to \ldots \to \nonumber \\
&& \qquad 
E^{k_1 + k_3 + h} N^{1 + k_2} W^{h + 1} \to
E^{k_1 + k_3 + h} N W^{h + k_2 + 1}.
\end{eqnarray}
It is obvious that these last transformations can also be applied 
when $p=0$ (in this case we have also $k_3 = 0$). Therefore all walks can be transformed into new ones 
of the form $E^p N W^q$. If $q=0$ the walk is directed, while for 
$q=1$ we can transform it into $E^p N E$ which is also directed.
For $q\ge 2$ we transform the walk into $E^p N^2 W^{q-1}$. 
Since $\omega(0)$ is not a nearest neighbour of the boundary there
is no obstruction to this transformation. 
If $q=2$, by means of an L1 move, we obtain a directed walk. 
For $q\ge 3$ we apply repeatedly the following sequence
\begin{eqnarray}
E^p N^2 W^{q-1} & \to  &E^p N^2 W^{q-2} S \to 
E^p N^2 W S W^{q-3} \to \nonumber  \\
&& E^{p+1} N W^{q-1} \to E^{p+1} N^2 W^{q-2},
\end{eqnarray}
obtaining a walk with $q=2$ and then a directed walk. Using lemma 1, 
the walk can be reduced to any rod.

Finally we consider case (d). If $k_2 = 0$, since $h>0$, we have 
$k_1 = 0$. Then using L0 moves we can deform the walk into 
$W^h N^{p+1} E^{k_3 + h}$. If $k_2 > 0$, 
using L0 moves we can modify the 
walk into $E^{k_1} N W^h N^{k_2 + p} E^{h + k_3}$.
Then, by applying repeatedly the transformation
\be
\omega \to E^{k_1 - 1} N W^h N E N^{k_2 + p - 1} E^{k_3 + h} 
\to  E^{k_1 - 1} N W^h N^{k_2 + p} E^{k_3 + h + 1},
\ee
we obtain a new walk of the form $N W^h N^{k_2 + p} E^{k_1 + k_3 + h}$
and then $W^h N^{k_2 + p+1}  E^{k_1 + k_3 + h}$. Thus the walk can always 
be deformed into a new one of the form $W^p N^l E^q$, which can be reduced to a
rod following the method applied to $E^p N W^q$.

Q.E.D.

\vskip 0.4truecm

It should be noticed that the hypothesis of the lemma 
is a necessary condition for its validity.
Indeed, consider a walk with $\omega(0) = (0,1)$, 
$\omega(N-2) = (-1,0)$, $\omega(N-1) = (0,0)$, $\omega(N) = (1,0)$. 
This walk cannot be deformed into a rod, since the endpoint is frozen.
This hypothesis is not required if one considers walks 
for which the first point is not fixed, allowing moves L1 also on the 
first step of the walk.

\vskip 0.4 truecm

\noindent
{\em Lemma 3}: Consider a walk $\omega$ in a two-dimensional 
strip of width $w\ge 2$.
Then $\omega $ can be reduced to a quotient-directed walk using the 
kink-kink bilocal algorithm.

\begin{figure}
\hspace*{1cm}
\begin{center}
\epsfxsize = 0.9\textwidth
\leavevmode\epsffile{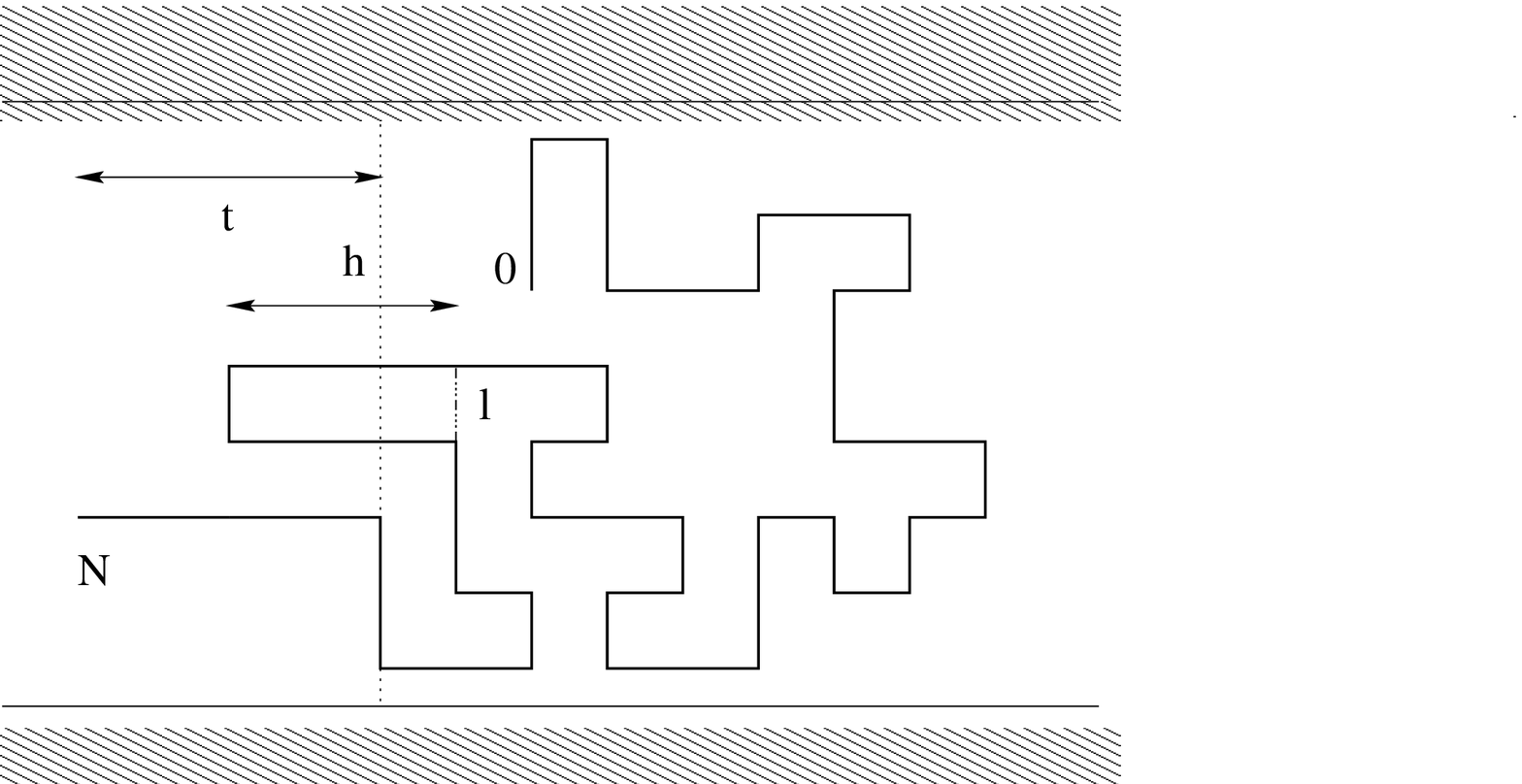}
\end{center}
\caption{The definition of $t$ when $|x_1-x_N| = 0$. 
The dotted line has equation $x=\overline{x}$, 
$h$ is the height of the tower and  $l$ its base.}
\label{fig_definitiont}
\end{figure}

\vskip 0.1truecm
\noindent Proof:

Let us first introduce a few notations.
To every walk $\omega$ (which is not a rod) with tower $T(l,h)$ we
associate a triple $(\omega,T(l,h),t)$. To define $t$, set $\overline\omega=
\omega/T(l,h)$ and consider $R[\overline\omega]$. Define
$R_1$ and $R_2$ as the two sides of $R[\overline\omega]$ which are
perpendicular to the boundary of the strip. Let us introduce
coordinates so that the $x$-axis is along the strip. If $R_1$ and
$R_2$ have equations $x=x_1$ and $x=x_2$ respectively, and
$(x_0,y_0)$, $(x_N,y_N)$ are the coordinates of $\omega(0)$ and
$\omega(N)$ respectively, it is not restrictive to assume
$|x_1-x_N|\leq |x_1-x_0|$ and $x_1\leq x_2$.
If $|x_1-x_N|\not=0$, set $t=0$. Otherwise find
the smallest $\overline x$ such that the line $x=\overline x$
contains a step of $\overline\omega$ or the starting point $ \omega (0)$
(such a line always exists: in the worst case the walk is a rod parallel to
the strip and $ \overline x = x_2 = x_0 $).
Then set $t=\overline x - x_1$. See Fig. \ref{fig_definitiont} for an
example.

The proof is by induction. The inductive step is the following:
given $(\omega,T(l,h),t)$ such that $ \overline\omega $ is not directed, the
tower is parallel to the strip,
all points of $T(l,h)$ lie on the $W$-side with respect to the base, and
$\omega(0)$ does not belong to the line $x=x_1 + t$, then there is a
sequence of moves such that: a. the new walk also has a tower parallel to the
strip, lying on the $W$-side with respect to the base; b. either $t$ increases
or $t$ remains constant but $h$, the height of the tower, increases.

To start the induction it is sufficient to
assume $h=0$ and choose as $l$ an arbitrary
step perpendicular to the boundary of the strip (it exists since 
$ \overline\omega $ is not a rod). At the end of the
inductive process we will obtain a new walk, which we will continue to name
$ \omega $, such that either $ \overline\omega $ is directed or
$ \omega (0) $ belongs to the line $ x = x_1 + t $. In the first case,
we have finished. The second case will be discussed
at the end.

Let us now prove the inductive step.
Let us first suppose that $x=x_1 + t$ contains a C-turn
$\overline\omega[k,m]$. Then, if the base of the tower
does not yet belong to $\overline\omega[k+1,m-1]$, use B22 moves
to move the tower so that the base is $\Delta\overline\omega(a)$,
$k+1\le a\le m -2$, and
the tower lies on the $W$-side of the base. It is easy to see that, thanks
to the definition of $t$, this is always possible. If
$\overline\omega[k,m]$ is unobstructed, then by applying L0
moves it is possible to modify $\overline\omega[k,m]$ in such a way that
the subwalk of $\omega$ connecting $\overline\omega(a-1)$ and
$\overline\omega(a+2)$ is a tower directed to W of height $h+1$.
In this way either $t$ increases by one or $t$ remains constant
but $h$ increases by one.

If $\overline\omega[k,m]$ is obstructed, there exists an unobstructed
C-turn $\overline\omega[i,j]$ (see theorem 1).
If the base of the tower does not belong to
$\overline\omega[i,j]$, reduce it to a kink, cut it and
increase the height of the tower by one.
If it contains the base of the tower, let us notice
that $\overline\omega[k,m]$ must have length at least 4 as it is obstructed. 
Therefore there exists a step belonging to
$\overline\omega[k+1,m-1]$ which does not belong to
$\overline\omega[i,j]$. Then move the tower on this link (this can be done
simply by B22 moves if this link is not adjacent to the base of the
tower, or by L0 moves if this is not the case) and at this point
reduce $\overline\omega[i,j]$ to a kink, and then increase the 
height of the tower by one, using B22 moves. In both cases, thanks
to the definition of $t$, the tower can always grow in the $W$-direction.
In this case $t$ remains constant, but $h$ increases by one.

Let us now suppose that the line $x=x_1+t$ does not contain any C-turn.
Since, by hypothesis, it does not contain $\omega(0)$, it can contain
only a subwalk $\overline\omega[i,j]$ such that 
either $j=N-2h$ or 
$\overline\omega[j,N-2h]$ is a line perpendicular to $R_1$, lying
on the W-side of $x=x_1+t$. If the base of the tower does not belong to
$\overline\omega[i,j]$, with an appropriate choice of $y$-direction 
we have $\omega[i,N] = S^{N-i-t} W^t$. Then, using L0 and L1 moves,
it is trivial to modify the walk into a new one such that 
$\omega[i,N] = W^{N-i}$. Therefore $t$ increases of at least $|j-i|$,
 proving the 
inductive step. If the base of the tower belongs to $\overline\omega[i,j]$, 
with an appropriate choice of $y$-direction
$\omega[i,N]$ is of the form 
$S^{k_1} W^{h} S E^{h} S^{k_2} W^t$ with $h>0$. Moreover,
if $x_l$ is the $x$-coordinate of $\omega(l)$, we have $x_l > \overline{x}$
for all $l < i$.
We will now distinguish two cases:
(1) $k_2>0$, (2) $k_2 = 0$. 

If $k_2 > 0$, using L0 moves, we can rewrite it as 
$S^{k_1 + k_2 - 1} W^{h} S E^{h} S W^t$. Then, using B22 moves followed 
by local transformations L0, we have
\be
S^q W^{h} S E^{h} S W^t\to S^q W^{h+t} S E^{h} S\to 
    S^q W^{h+t} S^2 E^{h}.
\ee
If $h > 1$ we can repeatedly modify the walk as follows:
\begin{eqnarray}
&& S^q W^{h+t} S^2 E^{h} \to S^q W^{h+t} S^2 E^{h-1} N \to
   S^q W^{h+t} S^2 E N E^{h-2} \to    \nonumber \\
&& S^q W^{h+t+1} S E^h \to S^q W^{h+t+1} S E^{h-1} S \to
   S^q W^{h+t+1} S^2 E^{h-1}.
\end{eqnarray}
Then we obtain a walk with the original form and $h=1$. But a walk 
$S^q W^p S^2 E$ can be modified into $S^q W^p S^2 W$, and then, by means
of L0 and L1 moves, into a rod $W^{p+q+3}$. Therefore in case (1)
the subwalk $\omega[i,N]$ can be transformed into a rod.

Let us now consider the case $k_2=0$. Since $h>0$, 
we should have $t=0$ so that 
$\omega[i,N]\to S^{k_1} W^{h} S E^{h}$.
If the endpoint of the walk does not belong to the boundary of the 
strip we can repeat the steps presented for case (1). 
If the endpoint belongs to the boundary and $k_1 > 0$, we can use
L0 moves to modify the walk into $S^{k_1 - 1} W^{h} S^2 E^{h}$, which
can be transformed into a rod as discussed in case (1). 
If $k_1 = 0$, since $w \ge 2$, we can apply the following 
transformations:
\noindent
if $ h = 1 $ then
\be
 \omega[i,N] = W S E \to W S W \to W^2 S \to W^3;
\ee

\noindent
if $ h = 2 $ then
\begin{eqnarray}
 \omega[i,N] & = & W^2 S E^2 \to N W S^2 E \to
 N W S^2 W \to \nonumber \\
 && N W^2 S^2 \to \cdots \to N W^4 \to \cdots \to W^5;
\end{eqnarray}

\noindent
if $ h > 2 $ then
\begin{eqnarray}
 \omega[i,N]& = & W^{h} S E^{h} \to N W S W^{h-2} S E^{h-1} \to
 N W^{h-1} S^2 E^{h-1} \to  \nonumber \\
 && N W^{h-1} S^2 E^{h-2} N \to N W^{h-1} S^2 E N E^{h-3} 
\to \nonumber \\
 && N W^{h} S E^{h-1} \to \cdots N W^{h} S^2 E^{h-2} \to
    \cdots \to NW^{2 h -3} S^2 E.
\end{eqnarray}
Then we can use the same move discussed in the case $h=2$, obtaining
$W^{2 h + 1}$.
Therefore in all cases we deform $\omega[i,N]$ into a rod. For the 
new walk the variable $t$ increases at least by $2 h + 1$ as required.

The proof of the inductive step has ended.
In this way we have shown that in a finite number of steps
we obtain a walk $\omega$ which 
is quotient-directed or is such that $x_0 = x_1 + t$. In the latter case, 
if $x_2=x_0$, $\overline\omega$ is L-shaped and thus directed. Thus,
we need only to study the case $x_2>x_0$. Again we will proceed by induction.
We assume that the walk $\omega$ has a tower parallel to the
strip which lies on the E-side with respect to the base. At the beginning
notice that $R_2$ must certainly contain a C-turn as it
contains none of the endpoints. Then move the tower on this
C-turn. This is always possible.
Then we show that as long as $x_2>x_0$ and the walk is not directed
we can modify it in such a way that the tower increases in height.
The argument is exactly identical to the one we have previously
discussed for the case in which a C-turn exists on the line $x=x_1+t$. 
Therefore, in a finite number of steps, we obtain a walk with 
$x_2 = x_0$. As we already discussed, this walk is quotient-directed.

Q.E.D.

\vskip 0.4truecm
It is now trivial to state the ergodicity theorem for the kink-kink 
bilocal algorithm which is a simple consequence of the lemmas proved above:

\vskip 0.4truecm

\noindent
{\em Theorem 3}: Consider a walk $\omega$ in a two-dimensional 
strip of width $w\ge 2$ and suppose that 
$ \omega (0) $ is not a nearest neighbour of a boundary of the strip. 
Then $\omega $ can be reduced to a given rod using the kink-kink bilocal 
algorithm.

\vskip 0.4truecm

The result we presented above applies only to the two-dimensional case. 
Indeed, the algorithm is not ergodic in three dimensions.
For instance, consider the walk $(N=18)$
\be 
\omega \equiv (-y)^2(x)(z)^2(-x)^2(-z)^3(x)^2(y)(z)^2(-x)(-y)^2. 
\ee
By direct enumeration, one can verify that it cannot be reduced to a rod.

The kink-kink bilocal algorithm can also be used to simulate ring polymers. 
It is enough to exclude the L1 moves. Such an algorithm was considered 
in \cite{Reiter_90,Jagodzinski-etal_92}. 
It is clear that this algorithm is not ergodic in three dimensions since 
it does not change the knot type of the ring.
We will now prove its ergodicity in a two-dimensional strip. 
We will need the following lemma:

\vskip 0.4truecm

\noindent
{\em Lemma 4}: In two dimensions, consider a closed walk $\omega$, 
with $N>4$ steps.
It contains an unobstructed C-turn $\omega[i,j]$ with 
$0\le i < j < N$.

\vskip 0.1truecm
\noindent Proof:
Consider $w=\omega[0,N-1]$. It is not directed, and thus, by theorem 1,
it contains an unobstructed C-turn $\omega[i,j]$ with
$0\le i < j < N$. Since $N>4$, it is also unobstructed in $\omega$.
Q.E.D.

\vskip 0.4truecm

We can now prove the ergodicity of the algorithm.

\vskip 0.4truecm

\noindent
{\em Theorem 4}: Consider a lattice point $x$ in a two-dimensional 
strip and the ensemble ${\cal R}_{x,N}$ of $N$-step self-avoiding 
polygons  such that
$\omega(0)=\omega(N) = x$. The kink-kink bilocal algorithm 
without L1 moves is an ergodic algorithm for ${\cal R}_{x,N}$.

\vskip 0.1truecm
\noindent Proof:
The proof is similar to that of Lemma 3. Consider a walk $\omega$ with 
tower $T(l,h)$ along the strip and let $\overline{\omega}=\omega/T(l,h)$.

We will now prove the following: if the length of $\overline{\omega}$
is larger than 4, we can modify the walk into a new one 
$\omega'$ with tower $T(l',h')$ parallel to the strip and $h' > h$.
Therefore, in a finite number of steps, $\overline{\omega}$ is 
reduced to a square of length 4 and $\omega$ is a rectangle of 
height one. It is trivial to show that all these rectangle can be 
modified one into the other ending the proof.

To prove the previous statement, first notice that at least 
one of the boundaries of the enveloping rectangle
$R[\overline{\omega}]$ perpendicular to the strip does not contain 
$\omega(0)$ and contains a C-turn $\overline{\omega}[k,m]$, $0\le k < m \le N$. 
Using B22 moves we first move the tower on a link $l \in
\overline{\omega}[k+1,m-1]$, adding kinks on top of this link,
outward with respect to $R[\overline{\omega}]$. 
If $\overline{\omega}[k,m]$ is unobstructed we can increase the height
of the tower using L0 moves, otherwise, by lemma 4, there exists an
unobstructed C-turn $\overline{\omega}[i,j]$ such that 
$\omega(0)\not\in \overline{\omega}[i+1,j-1]$. As we already
discussed in Theorem 3, such a C-turn
can be reduced to a kink which is then moved on top of the tower,
increasing its height. 

Q.E.D.

\subsection{Ergodicity for the extended reptation algorithm}

We will now prove the ergodicity in two dimensions of the 
extended reptation algorithm. We do not know whether this algorithm 
is ergodic for larger values of $d$.

\vskip 0.4truecm

\noindent
{\em Theorem 5}: The extended reptation algorithm is ergodic in $d=2$, 
for $w\equiv w_2 \ge 1 $.

\vskip 0.1 truecm

\noindent
Proof:
The proof is similar to that of Lemma 3. We assume we have a walk 
$\omega$ with a tower $T(l,h)$ parallel to the $x$-axis. Then we 
consider the enveloping rectangle $R[\overline{\omega}]$: $R_1$ and 
$R_2$ are the two sides of $R[\overline{\omega}]$ perpendicular to the 
boundary of the strip. We will now show that it is possible to 
deform the walk so that one of the endpoints belongs 
either to $R_1$ or $R_2$ (defined for the new walk).

The proof is by induction: given a walk $\omega$ such that none of the 
endpoints belongs to $R_1$ or $R_2$ and which has a tower 
$T(l,h)$ parallel to the $x$-axis, we can deform it into a new walk with tower 
$T(l',h+1)$ parallel to the $x$-axis. 
To prove this statement, notice that $R_1$  
must contain a C-turn $\overline{\omega}[i,j]$. 
If $l\not\in \overline{\omega}[i+1,j-1]$, using B22 moves, we can 
modify the walk so that it has a tower of height $h$ on a link 
belonging to $\omega[i+1,j-1]$ on the W-side with respect to $R_1$.
Since the walk has a C-turn, by theorem 1, it has an unobstructed C-turn. 
If $\overline{\omega}[i,j]$ is unobstructed, we can reduce it to a kink
obtaining a tower of height $h+1$. 
If $\overline{\omega}[k,l]$ with $l<i$ or $k>j$ is unobstructed,
we can reduce it to a kink and use B22 moves to move the kink on 
top of the tower increasing its height. We must then consider the 
special cases in which $\overline{\omega}[i,j]\cap \overline{\omega}[k,l]$
is not empty and $\overline{\omega}[i,j]$ is obstructed. 
It is easy to realize that one should have $l=i+2$ or $k=j-2$ and that 
they can be treated similarly. If $l=i+2$, since $|j-i|\ge 4$
($\overline{\omega}[i,j]$ is obstructed) we can modify the walk 
(if needed) so that the base of the tower does not belong to 
$\overline{\omega}[k,l]$. Then we reduce it to a kink, and move it on top of
the tower increasing its height. If $k=j-2$ we can proceed analogously.
We have thus proved the inductive step and therefore we have shown 
that in a finite number of steps we can obtain a new walk such that: 
(a) one of the endpoints belongs to $R_1$ or $R_2$;
(b) the base of the tower belongs to $R_1$. If one of the endpoints 
belongs to $R_2$, using reptation moves in the E-direction, we can reduce 
the walk to a rod. If none of the endpoints belongs to $R_2$, 
an endpoint belongs to $R_1$ and $R_2$ contains a C-turn 
$\overline{\omega}[i,j]$. Then, by means of B22 moves, one can move the 
tower on one of the links belonging to $\overline{\omega}[i,j]$, so that 
all points of the tower lie on the E-side with respect to the base. 
Then, consider the endpoint belonging to $R_1$.
Using reptation moves in the W-direction, we can reduce the walk to a rod.
Finally note that, if $w\ge 1$, the two rods $E^N$ and $W^N$
can be deformed one into the other. We have thus proved the ergodicity of 
the algorithm.

Q.E.D.

\vskip 0.4truecm
\noindent
It is easy to see that the moves that are added to the reptation algorithm
(L0 and B22) are necessary for the ergodicity of the 
algorithm. 
Indeed consider the walk with $N=22$:
\be
N^2 W^2 S^2 E S E^2 N^2 E^2 S E S^2 W^2 N^2.
\ee
It is easy to see that it cannot be deformed without using L0 moves.
The walk with $N=13$
\be
N W S^2 E^2 N E^2 S^2 W N
\ee
requires instead the B22 moves to be reduced to a rod.

\subsection{Ergodicity for the kink-end reptation algorithm}

In this Section we consider the kink-end reptation algorithm. We show
that this algorithm is ergodic in any cylinder in $d\ge 3$ and 
in the presence of a single surface (or in free space) in two 
dimensions. This algorithm is not ergodic in a two-dimensional
strip. However ergodicity is recovered by adding L0 moves.

\begin{figure}
\begin{center}
\epsfxsize = 0.9\textwidth
\leavevmode\epsffile{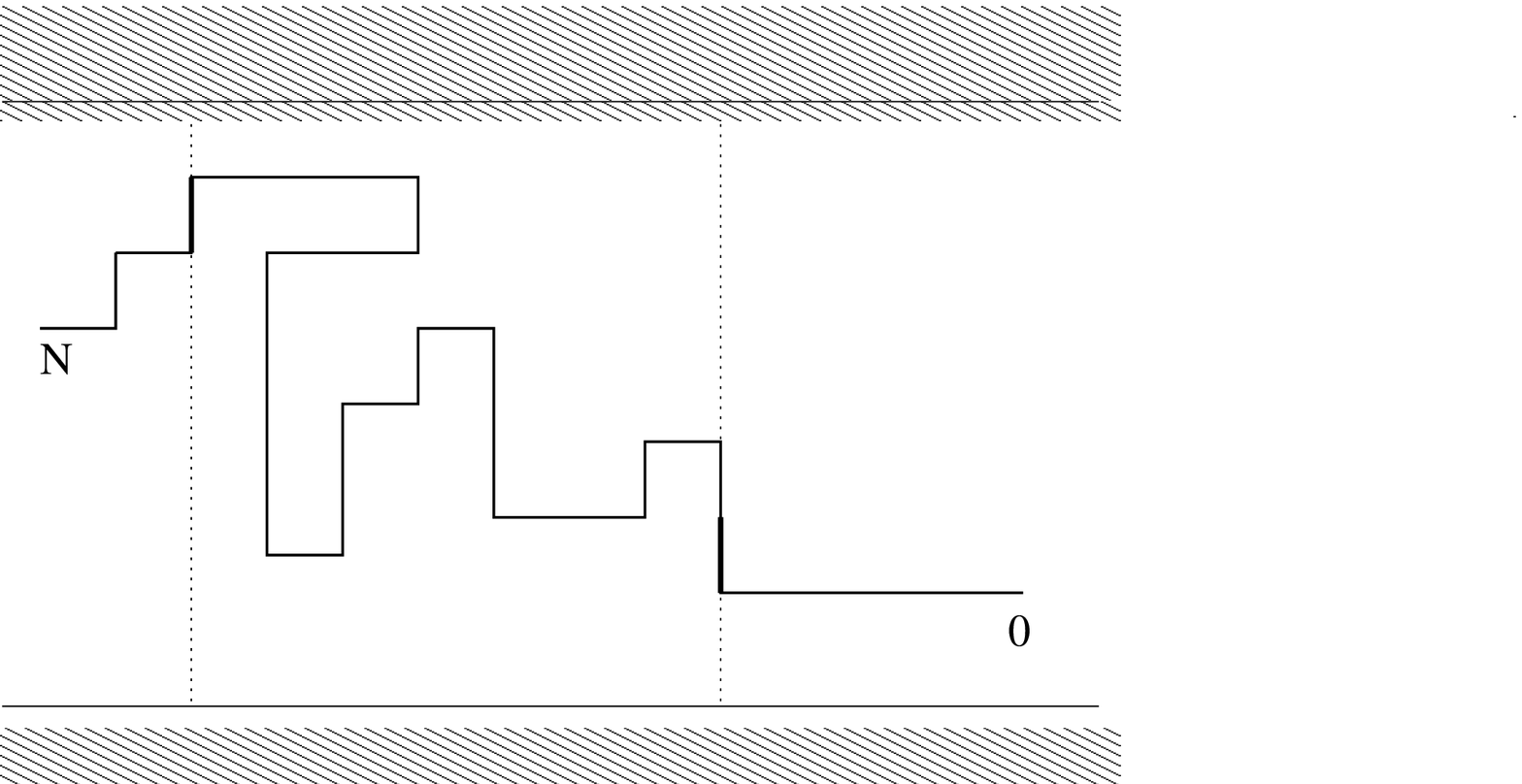}
\end{center}
\caption{With the definitions given in Theorem 6, the dotted vertical lines 
have equations $x=\overline{x}^\pm$. The step in boldface on the left 
is an $S$-step, while the step in boldface on the right 
is a step belonging to $x=\overline{x}^+$ that does not have an empty shadow
and is not an $S$-step.}
\label{fig_examplesTh5}
\end{figure}

\begin{figure}[t]
\begin{center}
\epsfxsize = 0.7\textwidth
\leavevmode\epsffile{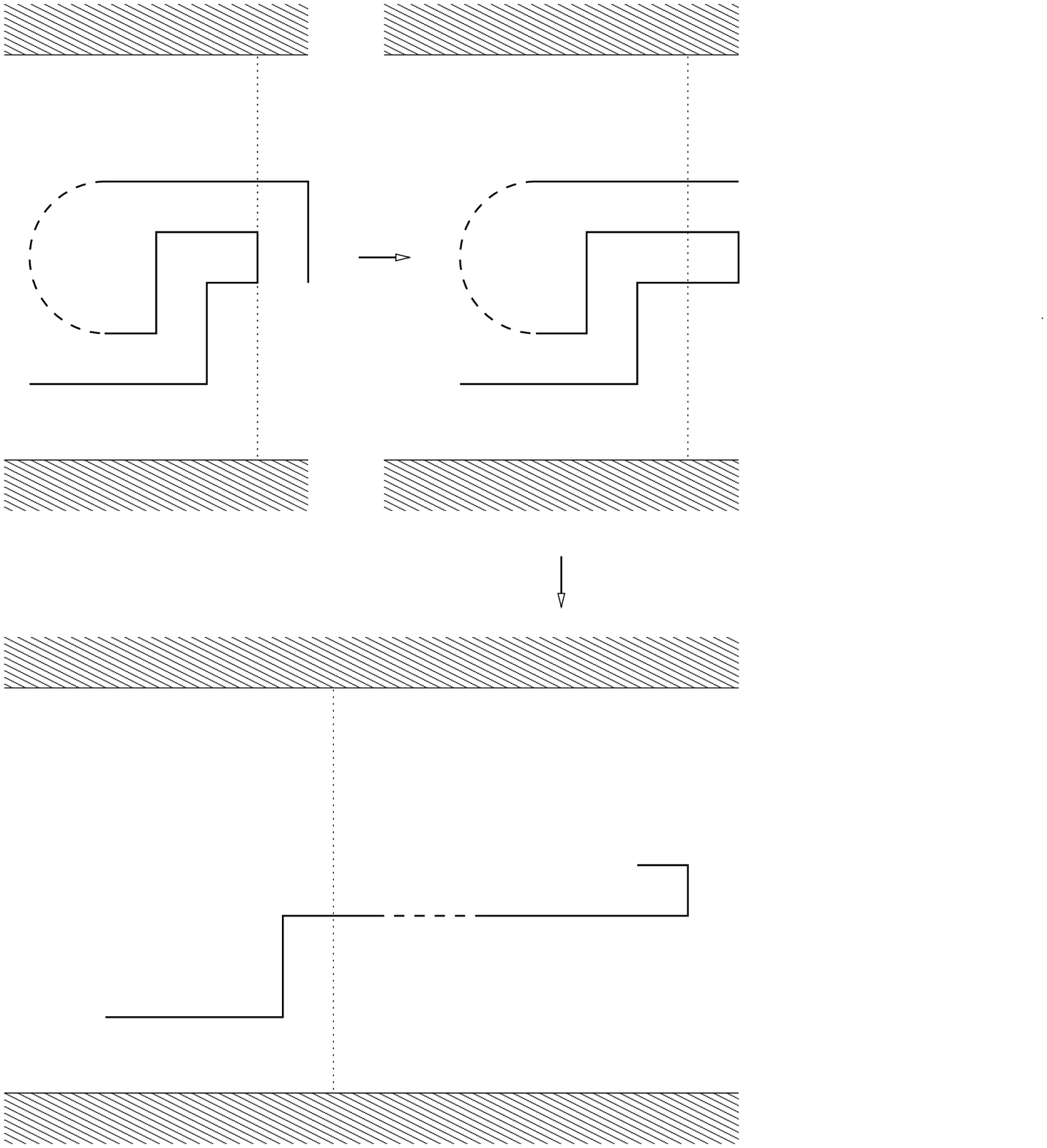}
\end{center}
\caption{The sequence of moves that generate a new walk $\omega'$ 
such that $\omega'[j,N-2]$ is a rod in the $(\pm x)$ direction.
The dashed line in the upper part of the figure 
indicates an arbitrary subwalk. The dotted line is the line of equation 
$x = \overline{x}^+$.}
\label{fig_inductionTh5}
\end{figure}

\vskip 0.4truecm
\noindent
{\em Theorem 6}: For $d \ge 3$, the kink-end reptation algorithm is ergodic 
for $w\equiv \min_i w_i\ge 1$ and $N \ge 3$.

\vskip 0.1 truecm
\noindent
Proof:  In order to prove the theorem let us introduce some useful definitions.
We say that a walk step $\Delta\omega(j)$, $1\le j\le N-4$,
is an {\em $S$-step} (see Fig. \ref{fig_examplesTh5})
if $\Delta\omega(j-1) = \Delta\omega(j+1)$ are 
directed in the $(\pm x)$-direction, while $\Delta\omega(j)$ is orthogonal
to them. Note that here we indicate by $x$ the first direction 
that, by definition,
is always infinite. Given a walk step $\Delta\omega(j)$ which is not in
the $x$-direction, we also define its positive and negative 
{\em shadow}. If $\hat{x}$ is the unit vector $(1,0,\ldots,0)$,
the positive (resp. negative) shadow is the set of lattice points 
$\{\omega(j) + n \hat{x},\omega(j+1) + n \hat{x}\}$, where 
$n$ is a positive (resp. negative) integer. We say that 
$\Delta\omega(j)$ has an {\em empty shadow} if its positive 
or negative shadow contains none of the walk sites $\omega(k)$, 
$0\le k \le N - 2$. To understand the relevance of this definition,
note that, if $\Delta\omega(j)$ has an empty shadow, 
then (see Fig. \ref{fig_inductionTh5})
we can perform end-kink moves by adding kinks on top of 
$\Delta\omega(j)$, obtaining eventually a new walk $\omega'$ such that
$\omega'[j,N-2]$ is a rod in the $x$-direction. Note that, 
by definition, the last two walk sites may belong to the shadow. 
However, they do not represent an obstruction for the end-kink
moves since the last two steps are deleted in the first 
iteration of the process.

The proof of the theorem is based on the following inductive step: given a walk
$\omega$ such that the subwalk $\omega[i,N-2]$, $0<i\le N-2$, is a rod 
in the $(\pm x)$-direction, we can deform it into a new walk 
$\omega'$ such that $\omega'[j,N-2]$, $j<i$, is also a rod in the 
$(\pm x)$-direction. Note that it is possible 
that, at the beginning of the induction process, $i = N-2$: 
it simply means that $\Delta\omega(N-3)$ is not in the 
$(\pm x)$-direction. This inductive step allows to prove that,
in a finite number of steps, any walk can be deformed
into $(\pm x)^{N-2}X$ where $X$ is the subwalk $\omega[N-2,N]$. 
Then, since $w\ge 1$ and $N\ge 3$, 
with an appropriate choice of the $y$-axis, we can 
deform it as follows:
\be 
(\pm x)^{N-2}X \to y (\pm x) (-y) (\pm x)^{N-2} \to (\pm x)^{N}.
\ee
Thus all walks can be deformed into a rod $(\pm x)^{N}$.
Finally, it is easy to show that the rod $(+x)^N$ can be deformed into
$(-x)^N$, proving the ergodicity of the algorithm.

To prove the inductive step, let us introduce coordinates 
$\omega(k) = (x_k,y_k,z_k, \ldots)$ and define 
$\overline{x}^-$ (resp. $\overline{x}^+$) as the smallest 
(resp. largest) value of $x$ such that there exists a walk site 
$\omega(k)$, $k < N - 2$, with $x_k=\overline{x}^{\pm}$ and 
$\Delta\omega(k)$ not in the $(\pm x)$-direction. 
If $\omega[0,N-2]$ is not a rod in the $(\pm x)$-direction,  
$\overline{x}^-$ and $\overline{x}^+$ certainly exist
although they may coincide.

Now consider the links belonging to the hyper-surfaces 
$x = \overline{x}^-$ and $x = \overline{x}^+$. Suppose that
one of them $\Delta\omega(l)$, $l< N -2$, is not 
an $S$-step. It is not restrictive to assume 
that it belongs to $x = \overline{x}^+$.  We will now show that there 
is $\Delta\omega(j)$, contained in $x = \overline{x}^+$ that has 
an empty shadow, with $j < i$. 
If $\Delta\omega(l)$ has an empty shadow, we can take $j = l$.
Since $l < N -2$, and all steps $\Delta\omega(i)$, \ldots 
$\Delta\omega(N-3)$ are in the $x$-direction, we have $j < i$. 
If $\Delta\omega(l)$ does not have an empty shadow, there are two
possibilities: (a) $\Delta\omega(l-1)$ is in the negative $x$-direction;
(b) $\Delta\omega(l+1)$ is in the positive $x$-direction.
We will consider only case (a) since case (b) is completely analogous.
In case (a) (see Fig. \ref{fig_examplesTh5}) 
we will now show that $\Delta\omega(l+1)$ has an empty shadow
and that $l+1< i$, so that we can take $j=l+1$. To prove this 
statement we will show the following: 
(a1) $l< N -3$;
(a2) $\Delta\omega(l+1)$ belongs to $x = \overline{x}^+$; 
(a3) $\Delta\omega(l+2)$ cannot be oriented in the positive $x$-direction.
>From (a2) and (a3) we see that $\Delta\omega(l+1)$ has an empty positive 
shadow, while (a2) and (a1) allow to conclude $l+1< i$ as required. 
To prove (a1), note that,
if (a1) were not true, we would have $l = N - 3$ and the walk
would be of the form $(-x)^{N-3} d_l X$, where $X\equiv \omega[N-2,N]$.
But this implies that $\Delta\omega(l)$ has an empty (negative) shadow, which 
is against the initial assumption.
To prove (a2), note that
$\Delta\omega(l+1)$ cannot be oriented in the negative $x$-direction;
otherwise, since $l<N-3$,
$\Delta\omega(l)$ would be an $S$-step. If it were directed in the positive 
$x$-direction, then the walk would be $(-x)^l d x^{N-l-3} X$ 
where $d$ is the direction of $\Delta\omega(l)$ and $X$ indicates the 
last two steps. But in this case $\Delta\omega(l)$ would have an empty
(negative) shadow. Therefore, (a2) is proved. 
If (a3) were not true, the walk would be 
$(-x)^l d_l d_{l+1} x^{N - l - 4} X$, where $d_l$ and $d_{l+1}$ 
are the directions of $\Delta\omega(l)$ and $\Delta\omega(l+1)$ 
respectively, and, therefore,  $\Delta\omega(l)$ would have an empty
(negative) shadow, against the initial hypothesis. 

If $\Delta\omega(j)$ has an empty shadow,
then we can deform the 
last steps of the walk as follows (we assume, without loss of generality,
to have a positive empty shadow):
\be
\omega[j,N]\to xy(-x) \ldots  \to x^2y(-x)^2 \ldots \to \ldots \to 
x^{N-2-j} y(-x),
\ee
where  $y$ is the direction of 
$\Delta\omega(j)$ (see Fig. \ref{fig_inductionTh5}). We have thus proved 
the inductive step.

Let us now suppose that all walk steps belonging to 
$x = \overline{x}^-$ and $x = \overline{x}^+$ are  
$S$-steps. In this case it is easy to convince oneself 
that, with an appropriate choice of axes, the walk has the 
form $x^j y X x^h Y$, where $j > 0$, $h > 0$, 
$X$ is the subwalk $\omega[j+1,N-h-2]$ 
and $Y$ is the configuration of the last two steps. 
Clearly, if $\omega(k)\in X$, then $\overline{x}^- < x_k \le \overline{x}^+$. 
If $\overline{x}^- =  \overline{x}^+$, $X$ is empty and 
$\omega = x^j y x^h Y$. Consider now $\Delta\omega(j)$ which is 
the only walk step\footnote{Note however that 
$\omega(N)$ may belong to the face $x=\overline{x}^-$. 
This is of no relevance for the following discussion since in an
end-kink move the last two steps are deleted.}
(as it can be seen from the explicit 
expressions above) belonging to the face $x = \overline{x}^-$. 
Since $w \ge 1$, it is possible to fix the positive $z$-direction 
in such a way that $\omega(j) + \widehat{z}$ 
($\widehat{z}$ is the unit vector in the positive $z$-direction) 
is inside the cylinder. Then we deform the walk as follows:
\be
\omega \to z x (-z) x^{j-1} \ldots \to
             (-x) z x^2 (-z) x^{j-1} \ldots \to 
              (-x)^{N-2} z x,
\ee
which can be reduced to a rod.
Thus in all cases we have proved the inductive step.

Q.E.D.

\vskip 0.4truecm

It is easy to see that this algorithm is not ergodic in a two-dimensional
strip of width $w\equiv w_2$.
Consider for instance the walks of the following form:
\be
( ( E N )^w E ( E S )^w E )^k E^2
\ee
with $ k \ge 1 $. The length of such a walk is $ N = (4w+2) k + 2 $.
It is easy to verify that they cannot be modified by the algorithm.

The previous theorem can be extended to two dimensions in free space,
or in the presence of a single boundary,  
i.e. for $w=\infty$. 

\vskip 0.4truecm
\noindent
{\em Theorem 7}: In two dimensions the kink-end reptation algorithm is ergodic 
for $N \ge 3$, in the presence of a single boundary or in free space.

\vskip 0.1 truecm
Proof: The proof follows the previous one. 
We should only modify the proof of the inductive step in the case 
in which only $S$-steps belong to the lines $x=\overline{x}^{-}$ 
and $x=\overline{x}^{+}$.
In this case, with a proper choice of axes,
the walk has the form 
\be
\omega = E^p N X E^h Y,
\label{formaomega}
\ee
$p\ge 1$, $h \ge 1$, where 
$X$ is the subwalk $\omega[p+1,N-h-2]$ which is 
contained between the lines $x=\overline{x}^\pm$ 
and $Y=\omega[N-2,N]$.
If $\omega(p)$ does not belong to the boundary, we can modify
the walk as follows:
\be
\omega = E^p N X E^h Y \to SENE^{p-1}N X E^h \to
      WSE^2NE^{p-1}\ldots \to W^{N-2}SE,
\label{eq21}
\ee
which can be reduced to a rod. 

We should finally discuss the case\footnote{If $p>1$, we can use the 
sequence of moves \reff{eq21} adding kinks ``above" the boundary.
The difficult case corresponds to $p=1$.}
in which $\omega[0,p]$ belongs to 
the boundary $y=0$. 
Let $\overline{y}$ be the largest value of 
$y$ such that there exists a walk step $\Delta\omega(l)$, 
$l < N -2$, belonging to the line $y = \overline{y}$. 
First we show that, if $y_k > \overline{y}$ ($y_k$ is the $y$-coordinate
of $\omega(k)$), then $k=N-1$ or $k=N$. Indeed, since $y_0=0$, if we had
$y_k > \overline{y}$ and $k\le N-2$, then $\omega[k-1,N-2]$ would point
$N$ which is in contrast with \reff{formaomega}. Now, consider the 
sites belonging to the line $y = \overline{y}$ 
and let $\omega(k)$ be the site with 
smallest $k$. Clearly $k \le i$. 
Note, moreover, that $\Delta\omega(k-1)$ is in the positive $y$ direction.
Indeed, it does not lie on $y = \overline{y}$, otherwise $\omega(k-1)$ 
would lie on this line (remember that $\omega(k)$ is the 
walk site with smallest $k$ belonging to this line). It is not oriented in the 
negative $y$-direction, 
otherwise, $y_{k-1} > \overline{y}$. For the same reason
$\Delta\omega(k+1)$ is not in the positive $y$-direction, unless 
$k+1=N-2$.
Then, we can use end-kink moves to put kinks on top of 
$\Delta\omega(k)$ in the positive $y$ direction. 
There is no obstruction to these moves, since the only possible 
walk sites that can
have a larger $y$ are the last two sites that are removed in the first
iteration of the process. 
In this way we modify the walk into a new one
(we keep calling it $\omega$) such that $\omega[k-1,N-2]$ is a rod
directed in the positive $y$-direction. 

Now let $\widetilde{x}$ be the largest $x$ such that there exists a walk
step $\Delta\omega(l)$, $l< N -2$, belonging to the line 
$x = \widetilde{x}$. Consider the steps belonging to this line and let
$j$ be the smallest integer such that $\Delta\omega(j)$ belongs to 
$x = \widetilde{x}$. Clearly $j \le k - 1 < k \le i$. 
Since $\omega[0,p]$, $p> 0$ is directed E and $\omega[i,N-2]$ is 
directed $N$, $\Delta\omega(j)$ has an empty positive shadow.
Therefore, by means of end-kink moves we modify the walk so that 
$\omega[j,N-2]$ is a rod in the positive direction. 
Since $ j < i$, we have proved the inductive step.

Q.E.D.
\vskip 0.4truecm

If one considers two dimensional strips, an ergodic algorithm can be obtained
adding L0 moves. 

\vskip 0.4truecm
\noindent
{\em Theorem 8}: In two dimensions the kink-end reptation algorithm 
with L0 moves is ergodic 
for $N \ge 3$ and $w_2 \ge 1$.

\vskip 0.1 truecm
Proof: The proof is identical to that of Theorem 6. We should only
change the proof of the inductive step for the case in which 
there are only $S$-steps on the lines $x=\overline{x}^{\pm}$. 
With a proper choice of axes the walk has the form \reff{formaomega}. Then,
using L0 moves, we modify the walk as follows
\be
\omega = E^{p-1} N E \ldots \to \ldots \to N E^p \ldots.
\ee
Then, using end-kink moves,
\be
\omega \to WNE^{p+1} \ldots \to \ldots \to
W^{N-2} N E,
\ee
which can be deformed into a rod.

Q.E.D.

\vskip 0.4truecm

Although the kink-end reptation algorithm is ergodic in two dimensions 
in the absence of confining surfaces, the proofs of the theorems 
indicate that ``staircase" sections of the walk (for instance 
sections of the form $\ldots ENENENEN \ldots$) will be changed very slowly 
by the algorithm. Therefore, in order to have an efficient implementation,
it is probably useful to include in all cases the L0 moves.

\section{Transition matrices}

In the previous Section we have discussed the ergodicity of the 
algorithms. Now, we discuss how to implement them 
in order to obtain the correct probability distribution. 
Here we will discuss how to use them to generate walks with 
uniform probability in the ensembles ${\cal E}_{x,N}$ or 
${\cal E}_N$.
Any other probability distribution can be obtained by adding a 
Metropolis test or a generalization thereof\footnote{
For simple lattice models of homopolymers and proteins, the 
Metropolis criterion should provide a plausible physical 
dynamics \cite{Taketomi-etal_75}. For a different point of view
see \cite{Zwanzig_97,Chan-Dill_98}.}.

\subsection{Kink-kink bilocal algorithm} 

We will begin by considering the kink-kink bilocal algorithm. 
Although not necessary to ensure the ergodicity of the algorithm,
we will also add the local L00 (crankshaft) moves. In order to
describe the algorithm it is important to classify the possible 
configurations of three successive links (see Fig. \ref{lowerg}):
\begin{figure}  
\centering
\vspace{0.3cm}
\epsfig{figure=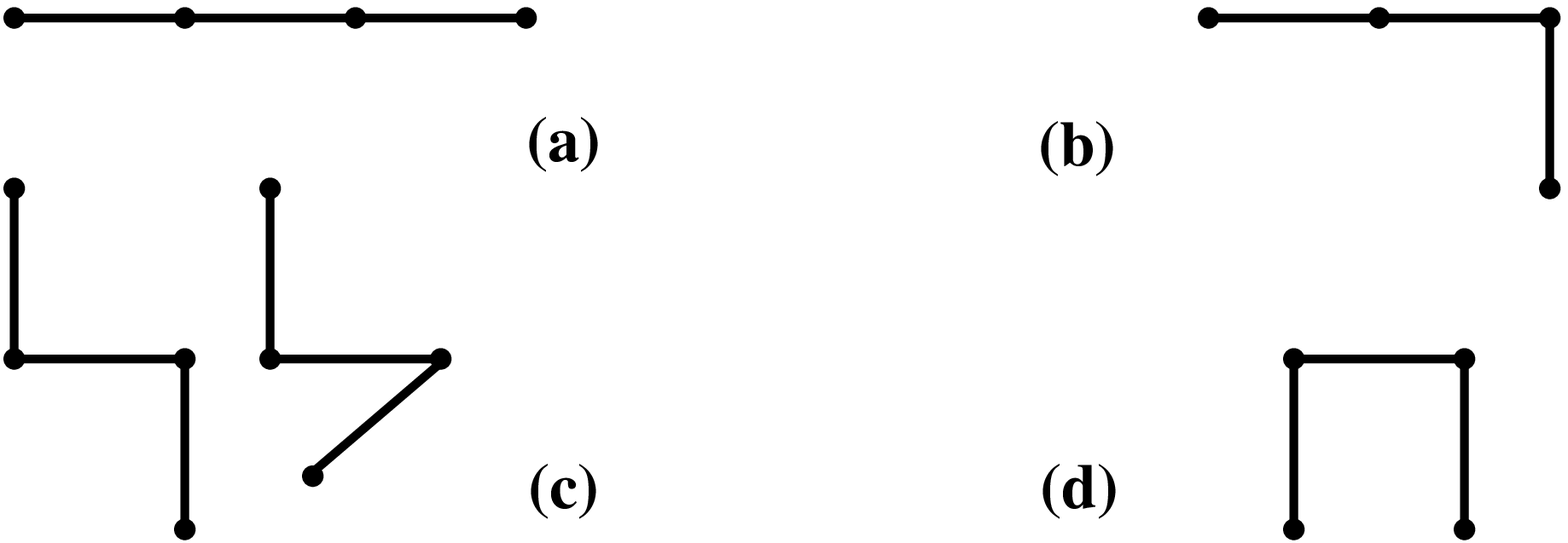,angle=0,width=0.95\linewidth}
\vspace{0.5cm}
\caption{
Configurations of three consecutive links: 
(a) configuration of type  $\textsf{I}$; (b)  configuration of type 
$\textsf{L}$; (c)  configuration of type $\textsf{S}$; 
(d)   configuration of type $\textsf{U}$.}
\label{lowerg}  
\end{figure}  
\begin{enumerate}
\item the bonds have the same direction (\textsf{I}
configuration);
\item two consecutive bonds have the same direction, while the third 
one is perpendicular to them (\textsf{L} config.);
\item the first and the third bond are perpendicular to the second one, and
they are either parallel or perpendicular to each other (\textsf{S}
config.);
\item the first and the third bond are perpendicular to the second one, and
they are antiparallel to each other (\textsf{U} config.).
\end{enumerate}

The algorithm works as follows:
\begin{itemize}
\item{Step 1.} Choose a random site $i$ of the current walk $\omega$,
$0\le i \le N$. If $i = N$, propose an L1 move and go to step 5.

\item{Step 2.} Determine the configuration of the subwalk 
$\omega[i-1,i+2]$. If $i = N-1$ we imagine adding a link 
$\Delta\omega(N)$ parallel to $\Delta\omega(N-1)$, so that 
the possible configurations are of type \textsf{L} and 
\textsf{I}. Analogously if $i = 0$, we imagine adding a 
link $\Delta\omega(-1)$ parallel to $\Delta\omega(0)$.

\item{Step 3.} Depending on the configuration of 
$\omega[i-1,i+2]$,  do the following:
\begin{enumerate}
\item{\textsf{I}}: with probability $1 - (2 d - 2) p(22)$ we perform a 
null transition. Otherwise we go the next step.
\item{\textsf{L}}: with probability $1 - (2 d - 3) p(22) - p(0)$ 
we perform a null transition, with probability $p(0)$ we propose an
L0 move and go to step 5. Otherwise we go to the next step.
\item{\textsf{S}}: with probability $1 - (2 d - 4) p(22) - 2 p(0)$ 
we perform a null transition, with probability $2 p(0)$ we try 
an L0 move: in this case there are two possibilities and we choose 
among them with equal probability and then go to step 5.
Otherwise we go to the next step.
\item{\textsf{U}}: 
with probability $(2 d - 3)p(00)$ we try an L00 move: there are 
$(2d - 3)$ possibilities which are chosen randomly; then we go to step 5. 
Otherwise we go to the next step.
\end{enumerate}

\item{Step 4.} 
Choose a second integer $j$ uniformly in the disjoint intervals, 
$-1\le j \le N$, $j\not = i-1,i, i+1$.
If $j = -1,N$ make a null transition. 
Then, depending on the configuration of $\omega[i-1,i+2]$, do the following:
\begin{itemize}
\item $\omega[{i-1,i+2}]$ is of type \textsf{I}, \textsf{S}, \textsf{L}:
if $j=0$ or $j = N - 1$, or if $\omega[j-1,j+2]$ is not of type 
\textsf{U} perform a null transition. Otherwise propose a B22 
move, cutting the kink $\omega[j-1,j+2]$ and adding it to 
$\omega[i,i+1]$ in one of the possible directions. Then go to the next step.
\item $\omega[{i-1,i+2}]$ is of type \textsf{U}:
according to the configuration of $\omega[j-1,j+2]$ (if $j=0,N-1$ imagine
adding links as before) do the following:
\begin{enumerate}
\item $\omega[j-1,j+2]$ is of type \textsf{I}: 
with probability $(2 d - 2)p(22)$ perform a B22 move: cut the kink
$\omega[i-1,i+2]$ and add it on top of $\omega[j,j+1]$ 
in a possible random direction, and then go to step 5. 
Otherwise perform a null transition.
\item $\omega[j-1,j+2]$ is of type \textsf{L}: 
with probability $(2 d - 3)p(22)$ perform a B22 move: cut the kink
$\omega[i-1,i+2]$ and add it  on top of $\omega[j,j+1]$ 
in a possible random direction, and then go to step 5. 
Otherwise perform a null transition.
\item $\omega[j-1,j+2]$ is of type \textsf{S}: 
with probability $(2 d - 4)p(22)$ perform a B22 move: cut the kink
$\omega[i-1,i+2]$ and add it on top of $\omega[j,j+1]$ 
in a possible random direction, and then go to step 5. 
Otherwise perform a null transition.
\item $\omega[j-1,j+2]$ is of type \textsf{U}: 
with probability $(2 d - 3)p(22)$ perform a B22 move: cut the kink
$\omega[i-1,i+2]$ and add it on top of $\omega[j,j+1]$ 
in a possible random direction, and then go to step 5; 
with probability $(2 d - 3)p(22)$ perform a B22 move: cut the kink
$\omega[j-1,j+2]$ and add it on top of $\omega[i,i+1]$ 
in a possible random direction, and then go to step 5. 
Otherwise perform a null transition.

\end{enumerate}
\end{itemize}
\item{Step 5.}: Check for self-avoidance. If the proposed new walk
is self-avoiding keep it, otherwise perform a null transition.
\end{itemize}

The algorithm we have presented depends on three probabilities 
$p(0)$, $p(00)$ and $p(22)$ that are the probabilities of 
an L0, L00 and B22 move respectively. It is easy to check that 
the algorithm satisfies detailed balance and thus the walks are 
generated with the correct probability distribution. We should now 
determine the single probabilities that must be such to satisfy 
the obvious constraint
\be
\sum_{\omega'} P(\omega\to\omega') = 1.
\label{sump1}
\ee
Considering the configurations \textsf{I}, \textsf{L}, and 
\textsf{S} we obtain the constraints
\begin{eqnarray}
(2d - 2)p(22) \leq 1, \\
(2d - 3)p(22) + p(0) \leq 1, \\
(2d - 4)p(22) + 2p(0) \leq 1 .
\end{eqnarray}
If $\omega[i-1,i+2]$ is of type \textsf{U}, we obtain, depending on the 
configuration of $\omega[j-1,j+1]$:
\begin{eqnarray}
(2d - 2) p(22) + (2d - 3)p(00) \leq 1, \\
(2d - 3) p(22) + (2d - 3)p(00) \leq 1, \\
(2d - 4) p(22) + (2d - 3)p(00) \leq 1, \\
2(2d - 3)p(22) + (2d - 3)p(00) \leq 1.
\end{eqnarray}
These conditions impose for $d \ge 2$:
\begin{eqnarray}
(4d - 6)p(22) + (2d - 3)p(00) \leq 1, \\
(2d - 4)p(22) + 2p(0) \leq 1.
\end{eqnarray}
A solution of these inequalities which maximizes $p(0)$ and $p(00)$ at
$p(22)$ fixed, is
\begin{eqnarray}
p(0) & = & \frac{1}{2}[1 - (2d - 4)p(22)], \\
p(00)  & = & \frac{1}{2d -3}[1 - (4d - 6)p(22)], \\
p(22) & \leq & \frac{1}{4d - 6}.
\end{eqnarray}
Since the L00 move is not necessary for the ergodicity of the algorithm,
while the B22 one is essential to ensure a fast dynamics, it is natural 
to require $p(22)$ to be maximal, even if this implies $p(00) = 0$. 
Then we obtain the following transition probabilities:
\begin{eqnarray}
p(0) & = & {d-1\over 4 d - 6}, \\
p(00)  & = &  0, \\
p(22) & = & \frac{1}{4d - 6}.
\end{eqnarray}
In two dimensions $p(0) = p(22) = 1/2$, while in three dimensions
$p(0) = 1/3$ and $p(22) = 1/6$.

\subsection{Extended reptation algorithm} 

This algorithm extends the standard reptation method. 
The reptation (or slithering-snake) algorithm has two different 
implementations. The first one, which satisfies detailed balance,
works as follows:
\begin{itemize}
\item Step 1. With probability 1/2 delete $\omega[N-1,N]$ and add a
new link at the beginning; otherwise delete $\omega[0,1]$ and add a new 
link at the end of the walk.

\item Step 2. Check if the new walk is self-avoiding. If it is keep it, 
otherwise perform a null transition.
\end{itemize}

A second version uses an additional flag which specifies which 
of $\omega(0)$ and $\omega(N)$ is the ``active" endpoint.
It works as follows:
\begin{itemize}
\item Step 1. Delete one bond at the ``active" endpoint and append a new 
one at the opposite end of the walk.

\item Step 2. If the new walk is self-avoiding keep it, otherwise 
stay with the old walk, and change the flag, switching the active endpoint.
\end{itemize}
This algorithm does not satisfy detailed balance, but it satisfies the 
stationarity condition generating the correct probability distribution.

The extended reptation algorithm consists in combining with 
non-zero probability the reptation algorithm and the kink-kink bilocal
algorithm.  More precisely the algorithm works as follows:
\begin{itemize}
\item Step 1. With probability $p$ perform a reptation move, 
with probability $1-p$ a kink-kink bilocal move, as specified in the 
previous section.
\end{itemize}
Note that in this algorithm the L1 moves are no longer needed. 
Therefore one can modify {\em Step 1.} of the kink-kink bilocal algorithm
choosing $i$ such that $0\le i \le N-1$. 
The probability $p$ is not fixed. It is only required that 
$0 < p < 1$ to ensure the ergodicity of the algorithm. 
It can therefore be tuned in order to obtain the best critical behaviour.

\subsection{Kink-end reptation algorithm}

The kink-end reptation algorithm uses kink-end and end-kink reptation 
moves (see Fig.~\ref{kink-end}).
We will present here two different implementations of the algorithm 
which, however, are expected to have the same 
critical behaviour.

Let us explain the first implementation. An iteration 
consists of the following steps:
\begin{itemize}
\item{Step 1.} Choose a random site $i$ of the current walk with 
$0\le i \le N - 3$.
\item{Step 2.} Try an end-kink move with probability $p(EK)$ or a
kink-end move with probability $p(KE) = 1 - p(EK)$. In the first case
delete the last two bonds of the walk and insert 
a kink on the bond $\Delta\omega(i)$ in one of the $(2 d - 2)$
possible orientations. In the second case,
if $i \not= 0$ and $\omega[i-1,i+2]$ is a kink,
remove it and attach two bonds at the end of the walk
in one of the $(2d - 1)^2$ possible ways; otherwise perform a
null transition.
\item {Step 3.} Check if the proposed walk is self-avoiding. If it is keep it,
otherwise make a null transition.
\end{itemize}
The transition matrix is given by
\be
P(\omega \rightarrow \omega') = \frac{1}{(N - 2)}p(T) \frac{1}{(2d - 2)}
\chi_{SAW}(\omega') \qquad
\textrm{if}\ T = \textrm{EK}
\ee
and
\be
P(\omega \rightarrow \omega') = \frac{1}{(N - 2)} p(T) \frac{1}{(2d -
1)^2}
\chi_{SAW}(\omega') \qquad \textrm{if}\ T = \textrm{KE}
\ee
for  $\omega \ne \omega'$, where $\chi_{SAW}$  is defined by
\be
\label{funcarsaw}
\chi_{SAW}(\omega') = \left \{ \begin{array}{ll}
1 &  \textrm{if $\omega'$ is self-avoiding} \\
0 & \textrm{otherwise.}
\end{array} \right.
\ee
Detailed balance requires 
\be
p(KE)(2d - 2) = p(EK)(2d -1)^2,
\ee
from which, using $p(EK) + p(KE) = 1$, we obtain
\be
\begin{array}{rcl}
p(KE) & = & \frac{(2d - 1)^2}{(2d - 1)^2 + (2d -2)}, \\
& & \\
p(EK) & = & \frac{2d - 2}{(2d - 1)^2 + (2d - 2)}, \\
\end{array}
\ee
In two and three dimensions we obtain explicitly
\be
\left. \begin{array}{lcl}
p(EK) & = & 2/11 \\
p(KE) & = & 9/11
\end{array} \right \} \textrm{in}\ d = 2; \qquad \qquad
\left. \begin{array}{lcl}
p(EK) & = & 4/29 \\
p(KE) & = & 25/29
\end{array} \right \} \textrm{in}\ d = 3.
\ee
The second implementation of the algorithm is similar to that of the
kink-kink bilocal algorithm. 
An iteration consists of the following steps:
\begin{itemize}
\item {Step 1.} Choose a random site of the current walk with 
$0\le i \le N - 3$.

\item{Step 2.} Determine the configuration of the subwalk
$\omega[i-1,i+2]$. If $i = 0$, we imagine adding a
link $\Delta\omega(-1)$ parallel to $\Delta\omega(0)$.

\item{Step 3.} Depending on the configuration of
$\omega[i-1,i+2]$,  do the following:
\begin{enumerate}
\item{\textsf{I}}: with probability $(2 d - 2) p(EK)$ 
perform an end-kink move, deleting the last two steps and 
adding a kink in one of the possible $(2 d - 2)$ directions;
otherwise perform a null transition.
\item{\textsf{L}}: with probability $(2 d - 3) p(EK)$ 
perform an end-kink move, deleting the last two steps and 
adding a kink in one of the possible $(2 d - 3)$ directions;
otherwise perform a null transition.
\item{\textsf{S}}: with probability $(2 d - 4) p(EK)$ 
perform an end-kink move, deleting the last two steps and 
adding a kink in one of the possible $(2 d - 4)$ directions;
otherwise perform a null transition.
\item{\textsf{U}}: with probability $(2 d - 3) p(EK)$ 
perform an end-kink move, deleting the last two steps and 
adding a kink in one of the possible $(2 d - 3)$ directions;
with probability $(2 d - 1)^2 p(KE)$ perform a kink-end move,
cutting the kink and adding randomly two links to the walk 
in random directions;
otherwise perform a null transition.
\end{enumerate}
\item Check whether the proposed new walk is self-avoiding. If it is
keep it, otherwise make a null transition.
\end{itemize}
The transition matrix is given by
\be
P(\omega \rightarrow \omega') = \frac{1}{(N - 2)}p(T) \chi_{SAW}(\omega')
\qquad
\textrm{where}\ T = EK\ \textrm{or}\ T = KE
\ee
for  $\omega \ne \omega'$, where $\chi_{SAW}$  is defined in
 Eq.(\ref{funcarsaw}). Detailed balance requires
$p(KE) = p(EK) \equiv p$, 
while Eq.~(\ref{sump1}) gives the following constraints:
\begin{eqnarray}
\label{eq1_1}
&& (2d - 2)p(EK) \leq 1, \\
\label{eq2_1}
&& (2d - 3)p(EK) \leq 1, \\
\label{eq4_1}
&& (2d - 4)p(EK) \leq 1, \\
\label{eq3_1} 
&& (2d - 1)^2 p(KE) + (2d - 3)p(EK) \leq 1. 
\end{eqnarray}
It follows
\begin{eqnarray}
p  & \leq & \frac{1}{2d - 2},\\
p  & \leq & \frac{1}{(2d - 3) + (2d - 1)^2}.
\label{pbetter}
\end{eqnarray}
For $d\ge 1$, the second inequality is the most restrictive one. 
Therefore the best choice
of $p$ corresponds to taking the equality in Eq. \reff{pbetter}.
In two and three dimensions we obtain 
$p =  1/10$ ($d = 2$) and $p = 1/28$ ($d = 3$).

\section{Dynamic critical behaviour}

In order to understand the efficiency of an algorithm 
one should analyze the autocorrelation time $\tau$. There are several
different definitions\footnote{We refer the reader interested in 
more precise and rigorous statements to 
\cite{Sokal_review,Madras-Slade_book}.}
for $\tau$: the {\em exponential} autocorrelation time $\tau_{exp}$
that controls the relaxation of the slowest mode in the system and 
the {\em integrated} autocorrelation time $\tau_{int,{\cal O}}$ that 
depends on the observable $\cal O$ one is considering and that 
controls the statistical errors on $\cal O$. For $N\to\infty$,
one expects a dynamic critical behaviour, i.e. $\tau\sim N^z$,
where the exponent $z$ may depend on which autocorrelation time one is 
considering. 

We now derive lower bounds on the exponent $z$ in the absence of 
interactions.
Let us consider global observables, like the squared end-to-end distance 
$R^2_e$ and the squared radius of gyration $R^2_g$. For bilocal 
algorithms we expect \cite{Sokal_review,Mandel_79}
$\tau_{int,{\cal O}}\gtapprox N^2$.
The basic assumption is that the slowest mode appearing in global
observables is associated to the relaxation of the 
squared radius of gyration $R^2_g$. Then, an estimate of $\tau$ can be obtained
as follows. At each elementary step, $R^2_g$
changes by a quantity of order $N^{2 \nu - 1}$. An independent configuration
is reached when the observable changes by one standard deviation 
$N^{2\nu}$. Assuming that the observable performs a random walk,
we obtain $\tau \sim (N^{2 \nu}/N^{2\nu - 1})^2 \sim N^2$. 
In practice the argument should provide only a lower bound\footnote{
In \cite{Caracciolo-etal_90} the following rigorous lower bound was 
obtained:
\be
\tau_{int,{\cal O}} \ge {2 {\rm var}({\cal O})\over C^2} - {1\over 2},
\ee
where ${\rm var}({\cal O})$ is the static variance of the observable 
${\cal O}$ and $C$ is the {\em maximum} change of ${\cal O}$ in a single 
Monte Carlo step. In the heuristic argument given above,
we have replaced $C$ by the {\em average} change of ${\cal O}$ 
in a single Monte Carlo step.} which we expect to be correct for all
global observables.

In particular, this should apply to the end-to-end distance $R^2_e$. Here, 
however, we should notice that our algorithms update the end-point
of the walk with very different frequencies. The extended reptation 
and the kink-end reptation change $\omega(N)$ every $O(1)$ iterations,
while the kink-kink bilocal algorithm updates $\omega(N)$ 
only every $O(1/N)$ iterations. Therefore, an additional factor of $N$ 
should be added for the kink-kink bilocal algorithm: in this case,
we expect $\tau\gtapprox N^3$.
It is interesting to notice that the kink-kink bilocal algorithm
behaves approximately as the algorithm of Reiter \cite{Reiter_90}:
indeed, also in this case, the endpoint is updated with frequency $1/N$.
The available numerical results for very short walks 
$(N\ltapprox 100)$ are in agreement with 
the bound given above: they indicate $\tau\sim N^3$
in two and three dimensions.

While Reiter's algorithm can be easily speeded up by increasing the 
frequency of the BEE and BKE moves, no improvement is possible
for the kink-kink bilocal algorithm. Indeed, in this case it makes no sense 
to increase the frequency of the L1 moves. Clearly, endpoint moves should be 
performed with the same frequency of the moves that change the 
site $\omega(N-1)$, otherwise they do not effectively change the endpoint
position. But $\omega(N-1)$ is updated by B22 and L0 moves with frequency 
$1/N$. Therefore L1 moves should be performed with the same frequency.

Let us now discuss in more detail the different implementations of 
the extended reptation and of the kink-end reptation algorithm.
For the extended reptation,
we should choose between the two different implementations of 
the reptation dynamics. If one considers ordinary random walks, 
it is obvious that 
a new walk is generated in $O(N^2)$ iterations of the first
algorithm and in exactly $N$ iterations of the second one.
Thus the second version is much more efficient than the first one.
For SAWs we do not expect such a big difference since the walk 
will move in a given direction only for a small number of steps 
($\approx 8$ in two dimensions, $\approx 14$ in three dimensions in the 
absence of interactions). 
Therefore we expect an improvement by a constant factor, and, indeed,
simulations \cite{Caracciolo-etal_2000} show that in three dimensions
the second implementation is 5--6 times faster that the first one.
\begin{table}
\begin{center}
\begin{tabular}{|l|rrrr|}
\hline
$d$ & $p(\textsf{I})$ &  $p(\textsf{L})$ & $p(\textsf{U})$
&  $p(\textsf{S})$ \\
\hline
2 & 0.152 & 0.481 & 0.108 & 0.259 \\
3 & 0.051 & 0.356 & 0.102 & 0.491 \\
\hline
\end{tabular}
\end{center}
\caption{Probabilities of the different configurations of three
links.}
\label{effprop}
\end{table}
In the extended reptation we should also fix the parameter $p$. From the 
discussion given above, it is clear that we must have 
$p > 0$ as $N\to\infty$, otherwise the motion of the endpoints slows 
down the dynamics. To fix its specific value, we may compare $p$
to the probabilities of proposing local and bilocal moves
in the kink-kink bilocal algorithm.
Assuming that the probability of occuring of a $\textsf I$, 
$\textsf L$, $\textsf U$, and $\textsf{S}$ configurations is 
independent of the position of the walk site --- it should be approximately true
for large values of $N$ --- the probability $p_b$ of a bilocal move
B22 and the probability $p_l$ of a local move L0 are given by
\begin{eqnarray}
p_b &=& 2 p(22) p(\textsf{U})
   \left[ (2 d - 2) p(\textsf{I}) +
          (2 d - 3) (p(\textsf{L}) + p(\textsf{U})) +
          (2 d - 4) p(\textsf{S}) \right],
\\
p_{l} &=& p(0) ( p(\textsf{L}) + 2 p(\textsf{S}) ),
\end{eqnarray}
To have a quantitative prediction we should know $p(\textsf{U})$,
$p(\textsf{I})$, $p(\textsf{L})$, and $p(\textsf{S})$. If we were considering
non-reversal random walk we would have 
\begin{eqnarray}
p(\textsf{U}) = {2 (d - 1)\over (2 d - 1)^2} \\
p(\textsf{I}) = {1\over (2 d - 1)^2} \\
p(\textsf{L}) = {4 (d - 1)\over (2 d - 1)^2} \\
p(\textsf{S}) = {2 (d - 1)(2 d - 3)\over (2 d - 1)^2} 
\end{eqnarray}
so that 
\begin{eqnarray}
p_b &=& {16 (d-1)^4\over (2d-3) (2d-1)^4},
\\
p_l &=& {4 (d-1)^3\over (2d-3) (2d-1)^2}.
\end{eqnarray}
In two and three dimensions we obtain 
$p_b\approx 0.1975$, $p_l \approx 0.4444$, and $p_b\approx 0.1365$, 
$p_l \approx 0.4267$ respectively.
For SAWs the probabilities can be computed by means of a short 
Monte Carlo simulation. The results are reported in Table \ref{effprop}.
In two dimensions we obtain $p_b = 0.096$, $p_{l} = 0.500$, while in
three dimensions  $p_b = 0.087$, $p_{l} = 0.446$. Thus, in the extended
reptation algorithm, B22 moves are proposed with probability 
$\approx 0.1 (1-p)$,  while reptation moves are proposed with 
probability $p$. If one wants to balance these two types of moves 
--- this is reasonable of one wants a physical kinetics --- one should choose 
$p\sim 0.1$. On the other hand, it is clear that reptation moves are more 
relevant than B22 moves. Indeed, reptation moves are essential for the 
motion of the endpoint, while B22 moves are required only to avoid the 
trapping of the endpoints. Thus we expect the algorithm to be more 
efficient for larger values of $p$. The simulations \cite{Caracciolo-etal_2000}
indicate that the fastest dynamics is obtained for 
$0.5\ltapprox p \ltapprox 0.9$.

Finally let us consider the kink-end reptation algorithm. We presented two 
different versions and we discuss now their relative efficiency. 
The first implementation chooses the move without checking the nearby bonds.
A deformation is always proposed but it may immediately fail because it 
does not respect self-avoidance when one considers the neighbours of the 
chosen bond. The second algorithm is more careful: the move is chosen
after considering the position of the nearby bonds. However, with 
a finite probability, it performs a null transition. In order to compare 
correctly the two implementations we should therefore compute:
\begin{itemize}
\item[(a)] for the first algorithm, the probability of bilocal moves
that do not fail after checking the position of the two nearby bonds;
\item[(b)] for the second algorithm, the probability of proposing a 
bilocal move.
\end{itemize}
For the first algorithm an easy computation gives
\begin{eqnarray}
p_{kink-end} &=& p({\textsf U}) p(KE) \\
p_{end-kink} &=& {p(EK)\over 2(d-1)}
   \left[ (2 d - 2) p(\textsf{I}) +
          (2 d - 3) (p(\textsf{L}) + p(\textsf{U})) +
          (2 d - 4) p(\textsf{S}) \right],
\end{eqnarray}
while for the second we have
\begin{eqnarray}
p_{kink-end} &=& p({\textsf U}) p (2d-1)^2 \\
p_{end-kink} &=& {p}\
   \left[ (2 d - 2) p(\textsf{I}) +
          (2 d - 3) (p(\textsf{L}) + p(\textsf{U})) +
          (2 d - 4) p(\textsf{S}) \right].
\end{eqnarray}
Therefore the first algorithm 
is equivalent to the second one with probability
\be
p = {p(KE)\over (2d-1)^2} = 
    {p(EK)\over (2 d - 2)} =\, {1\over (2 d - 2) + (2 d - 1)^2},
\ee
i.e. $p = 1/11$, $1/29$ in two and three dimensions respectively. 
This should be compared with the optimal value of $p$,
$p=1/10$ and $p = 1/28$ for $d=2,3$. Thus the second algorithm is more 
efficient than the first one, as it should be expected since the second one
chooses the proposed move more carefully. However the improvement in 
efficiency is small, approximately 10\% in two dimensions and only
3\% in three dimensions.

\section*{Acknowledgments}
We thank Alan Sokal for many useful comments.

\end{document}